\documentclass[pdflatex, 11pt]{article}
\usepackage{mathrsfs}
\usepackage{amssymb}
\usepackage{amsmath}
\usepackage{ascmac}
\usepackage{amsthm}
\usepackage[dvipdfmx]{graphicx}
\usepackage{booktabs}
\usepackage{setspace}
\usepackage{natbib} 
\usepackage{color}
\usepackage{float}
\usepackage{mathtools}
\usepackage{geometry}

\newcommand{\diag}{\text{diag}}

\newcommand{\vb}{\mathbf{b}}

\newcommand{\vd}{\mathbf{d}}

\newcommand{\vm}{\mathbf{m}}

\newcommand{\vr}{\mathbf{r}}

\newcommand{\vu}{\mathbf{u}}

\newcommand{\vx}{\mathbf{x}}
\newcommand{\vy}{\mathbf{y}}

\newcommand{\zero}{\mathbf{0}}

\newcommand{\mB}{\mathbf{B}}

\newcommand{\mI}{\mathbf{I}}

\newcommand{\mV}{\mathbf{V}}

\newcommand{\eps}{\epsilon}
\newcommand{\lam}{\lambda}

\newcommand{\vbeta}{\text{\boldmath{$\beta$}}}

\newcommand{\vlam}{\text{\boldmath{$\lambda$}}}

\newcommand{\mLam}{\mathbf{\Lambda}}
\newcommand{\mOmega}{\mathbf{\Omega}}
\newcommand{\mPhi}{\mathbf{\Phi}}

\newcommand{\M}{\mathcal{M}}

\begin{document}

\title{
	\textbf{Bayesian factor zero-inflated Poisson model for multiple grouped count data}
}
\date{}
\author{}

\maketitle

\maketitle

\vspace{-1.5cm}
\begin{center}
Genya Kobayashi$^{1}$\footnote{Author of correspondance: \texttt{gkobayashi@meiji.ac.jp}} and Yuta Yamauchi$^{2}$
\end{center}

\noindent
$^1$School of Commerce, Meiji University\\
$^2$Department of Economics, Nagoya University

\medskip
\begin{center}
{\bf  Abstract}
\end{center}
This paper proposes a computationally efficient Bayesian factor model for multiple grouped count data. 
Adopting the link function approach, the proposed model can capture the association within and between the at-risk probabilities and Poisson counts over multiple dimensions. 
The likelihood function for the grouped count data consists of the differences of the cumulative distribution functions evaluated at the endpoints of the groups, defining the probabilities of each data point falling in the groups. 
The combination of the data augmentation of underlying counts, the P\'{o}lya-Gamma augmentation to approximate the Poisson distribution, and parameter expansion for the factor components is used to facilitate posterior computing. 
The efficacy of the proposed factor model is demonstrated using the simulated data and real data on the involvement of youths in the nineteen illegal activities. 
\vspace{-0cm}


\section{Introduction}
Zero-inflation is a prevalent issue in the statistical analysis of count data in various applications, such as epidemiology, health services research, and social studies.
Several well-developed statistical methods exist for analysing zero-inflated count data, with the zero-inflated model \citep{Lambert92} being one of the commonly used approaches. 
See, for example, \cite{Neelon16} for a review.

Another significant challenge in the count data analysis arises from the occurrence of `grouped counts'. 
Instead of actual counts, grouped count data provide frequencies of individuals for predefined ordinal groups. 
Grouping occurs due to various factors, such as the sensitivity of the data topic and cognitive burden experienced by interviewees \citep{Fu18}.  
For example, in our real data analysis, the frequencies of involvement in illegal activities are reported in categories such as `never', `once', `twice', `between three and five times', `between six and ten times', `between eleven and fifty times' and `over fifty times', instead of the exact frequencies.  

Although there exists a body of studies analysing grouped continuous data, especially in the context of income data analysis
\citep[see, e.g., ][]{Kobayashi22, Kobayashi23}, the statistical analysis of grouped count data has much less attention, though grouped count data frequently arises,  especially in applied social science. 
To our knowledge, \cite{McG17} is the only study introducing the model for the grouped zero-inflated count data.  
\cite{McG17} employs the likelihood function of an ordinal response model where the likelihood contribution of each group is expressed by the difference between the cumulative distribution function of a discrete probability distribution evaluated at the endpoints of the group. 
These differences define the probabilities of the data points falling into the groups. 
However, when zero-inflation is high, an analysis of zero-inflated grouped count data using a univariate model can be distorted by the severe scarcity of information due to grouping and zero-inflation. 
If the data include multiple count responses, leveraging shared information among them by analysing them jointly considering a multivariate structure rather than treating them independently would be beneficial.

In addition, there has also been a growing demand for the joint analysis of multiple count data of which some or all dimensions are zero-inflated \citep[see, e.g.,][]{berry2020bayesian}. 
However, unlike continuous distributions such as the normal, developing and implementing a multivariate count model is generally cumbersome, especially when the multivariate counts are zero-inflated, as in the recent study of \cite{Liu15}.  

Factor analysis stands out as a common approach to analysing multivariate count data in a parsimonious and computationally convenient manner. 
To introduce a factor structure into count data analysis, a link function is commonly used to model a latent linear predictor incorporating latent factors and covariates \citep{Wedel03}. 
As an alternative approach, \cite{Larsson20} introduced a distinct type of factor model for discrete data, differing from classical count factor models, which is based on a dependent Poisson model  \citep[see, e.g.][]{Karlis03}. 
Some previous research exists on the factor models of zero-inflated count data, such as \cite{Neelon17} and \cite{Xu21}. 
\cite{Neelon17} introduced the factor structure into the at-risk probability and the mean count using the multiplicative function of the latent factor and regression components. 
\cite{Xu21} used the link-function approach to connect the zero-inflated count and latent linear predictor with factors.
The fundamental difference between our approach and the previous approaches lies in the flexibility of the factor structure. 
Due to its multiplicative structure, the factor model studied in \cite{Neelon17} permits only positive factors.
\cite{Xu21} employed the common latent linear predictor for both the at-risk probability and the mean count, which results in a restrictive correlation structure.

Based on the preceding, we propose the zero-inflated Poisson model with a flexible latent factor structure for multiple grouped count data. 
For modelling grouped count in each dimension, we follow \cite{McG17} and introduce the likelihood function for an ordinal model described above. 
To introduce the association within and between the at-risk and Poisson parts over different dimensions, we introduce the individual-specific latent factors with the dimension-specific factor loadings for the at-risk and Poisson parts. 
To facilitate posterior computation, we employ the P\'{o}lya-Gamma (PG) mixture representation of \cite{Polson13}. 
Since our model is Poisson-based, following \cite{HIS21}, we approximate the Poisson model by the negative binomial model and apply PG data augmentation. 
This augmentation enables us to carry out an efficient Gibbs sampling.
Moreover, for efficient sampling, we also borrow the idea of the parameter expansion technique of \cite{GD09} for the factor components, but without the positive lower triangular constraints. 
The MCMC draws of the unidentified working parameters are post-processed using the algorithm of \cite{Papa}. 
While achieving a stable sampling of the factor components in the low layer of the hierarchical model may seem challenging, our sampling method works well, as illustrated in the real data analysis where the counts are highly zero-inflated and highly coarsened into groups.

The remainder of this paper is organised as follows. 
Section~\ref{sec:method} introduces the proposed factor model for zero-inflated grouped counts. 
Then, the MCMC algorithm for the posterior inference is provided by applying the PG augmentation, data augmentation of the underlying counts, and parameter expansion. 
We also describe the post-processing for producing identified MCMC draws. 
The efficacy of the joint modelling through the latent factors is demonstrated by using the simulated data in Section~\ref{sec:sim} and real data in Section~\ref{sec:real}. 
Specifically, Section~\ref{sec:real} analyses the grouped count data of National Longitudinal Study of Youths 1979 (NLSY79) on the illegal activities by youths. 
Finally, Section~\ref{sec:conc} provides some conclusion and discussion.

\section{Method}\label{sec:method}
\subsection{Model}
Let $\vy_i=(y_{i1},\dots,y_{iJ})'$ denote the $J$ dimensional vector of the response variables. 
Each element of $\vy_i$ consists of zero-inflated grouped count data. 
Let $\vy_i^*=(y_{i1}^*,\dots,y_{iJ}^*)'$ denote the vector of the latent count data, and each element of $\vy_i^*$ is assumed to follow the zero-inflated Poisson distribution (ZIP) model expressed as
\[
y_{ij}^*\sim (1-\pi_{ij}) I(z_{ij}=0, y^*_{ij}=0) + \pi_{ij} Po(\mu_{ij})I(z_{ij}=1),\quad i=1,\dots,N,\quad j=1,\dots,J. 
\]
where $Po(\mu)$ denotes the Poisson distribution with the mean parameter $\mu$, $z_{ij}$ is the latent binary indicator such that $z_{ij}=1$ with probability $\pi_{ij}$ and $z_{ij}=0$ otherwise. 
If $z_{ij}=0$, the latent count is equal to structurally zero with probability and otherwise follows the Poisson distribution (at-risk). 
Given a known grouping mechanism $c$, $y_{ij}$ is observed as $c(y_{ij}^*)$. 
Generally, $c$  is in the form of
\begin{equation}\label{eqn:grouping}
y_{ij} = g \quad \text{iff}\quad \kappa_{g} \leq y_{ij}^*< \kappa_{g+1},\quad g=0,\dots,G-1,
\end{equation}
where $\kappa_g$'s define the thresholds of the ordinal groups \citep[see for example,][]{McG17}. 
Typically, $\kappa_g=0$ and $\kappa_G=\infty$. 
We utilise this data augmentation form for the posterior computation. 

The at-risk probability $\pi_{ij}=\Pr(z_{ij}=1)$  is modelled using the logistic model given by
\[
\pi_{ij} = \frac{\exp(\eta_{1ij})}{1+\exp(\eta_{1ij})},\quad i=1,\dots,N,\quad j=1,\dots,J. 
\]
The Poisson mean is modelled through the log-link function $\mu_{ij} = \exp(\eta_{2ij})$.

In order to connect the $2\times J$ responses, the common latent factor is introduced to the linear predictor in such a way that
\[
\eta_{hij} = \vx_{ij}'\vbeta_{hj} + \vu_i'\vlam_{hj}, \quad h=1,2, 
\]
where $\vx_{ij}$ is the $P\times1$ vector of covariates with the associated coefficient $\vbeta_{hj}$, $\vu_i=(u_{i1},\dots,u_{iK})'$ is the $K\times1$ common latent factor and $\vlam_{hj}=(\lam_{hj1},\dots,\lam_{hjK})'$ is the corresponding factor loading for $h=1,2$ and $j=1,\dots,J$. 

In order to facilitate the posterior computation and identification, we follow \citet{HIS21} to approximate  the Poisson model by the negative binomial model and apply the P\'{o}lya-Gamma (PG) mixture of \citet{Polson13}.
It is well known that the negative binomial distribution has the following mixture representation:
\[
Y\sim Po(\eps e^\eta),\quad \eps\sim Ga(r,r), 
\]
where $Ga(a,b)$ denotes the gamma distribution with the mean $a/b$. 
The marginal probability function of $Y$ is given by
\[
p(y)=\frac{\Gamma(y+r)}{\Gamma(r)y!}\frac{(e^\eta/r)^y}{(1+e^\eta/r)^{y+r}} = \frac{\Gamma(y+r)}{\Gamma(r)y!}\frac{(e^\psi)^y}{(1+e^\psi)^{y+r}},
\]
where $\psi=\eta-\log r$. 
The Poisson distribution is obtained in the limit of $r\rightarrow \infty$. 
Therefore, for a sufficiently large $r$, we can apply the P\'{o}lya-Gamma (PG) mixture representation to this approximate Poisson model:
\[
\frac{(e^\psi)^a}{(1+e^\psi)^b}=2^{-b}e^{\kappa\psi}\int_0^\infty e^{-\omega\psi^2/2}p(\omega|b,0)\text{d}\omega,
\]
where $a=y$, $b=y+r$, $\kappa=a-b/2$ and $\omega$ follows the PG distribution $PG(b,0)$ with the density $p(\omega|b,0)$.

Collecting the $2J$ terms of the PG mixture, the contribution of the $i$th individual to the augmented likelihood function conditionally on $\omega_{1ij}$, $\omega_{2ij}$,  $y_{ij}^*$ and $\prod_{j=1}^Jz_{ij}=1$ is proportional to
\begin{eqnarray}\label{eqn:Li}
&&
\begin{split}
&\prod_{j=1}^J 
    \exp\left\{-\frac{\omega_{1ij}}{2}(\vx_{i}'\vbeta_{1j}+\vu_i'\vlam_{1j})^2
    +\kappa_{1ij}(\vx_{i}'\vbeta_{1j}+\vu_i'\vlam_{1j})\right\}\\
&\quad\times\exp\left\{-\frac{\omega_{2ij}}{2}(\vx_{i}'\vbeta_{2j}+\vu_i'\vlam_{2j}-\log r)^2+\kappa_{2ij}(\vx_{i}'\vbeta_{2j}+\vu_i'\vlam_{2j}-\log r)\right\}
\end{split}\\
&&\propto\exp\left\{-\frac{1}{2}(\vd_i-\vbeta\vx_i-\mLam\vu_i - \vr)'\mOmega_i(\vd_i-\vbeta\vx_i-\mLam\vu_i  -\vr)\right\}\nonumber
\end{eqnarray}
where $\kappa_{1ij}=z_{ij}-1/2$, $\kappa_{2ij}=(y_{ij}^*-r)/2$, $\vr = (\underbrace{0,\dots,0}_{J},
\underbrace{\log r,\dots,\log  r}_{J})'$, 
$\vd_i=(d_{1i1},\dots,d_{1iJ},d_{2i1},\dots,d_{2iJ})'$ ($2J\times1$ vector),
$d_{1ij}=(z_{ij}-1/2)/\omega_{1ij}$, $d_{2ij}=(y_{ij}^*-r_j)/(2\omega_{2ij})$,  $\vbeta$ is the $2J\times p$ matrix such that $\vbeta'=(\vbeta_{11},\dots,\vbeta_{1J},\vbeta_{21},\dots,\vbeta_{2J})$, 
$\mOmega_i=\diag(\omega_{1i1},\dots,\omega_{1iJ},\omega_{2i1},\dots,\omega_{2iJ})$, $\mLam$ is the $2J\times K$ matrix such that $\mLam' = \left[\vlam_{11},\dots,\vlam_{1J},\vlam_{21},\dots,\vlam_{2J}\right]$. 
Conditionally on $\omega_{hij}$ and $\vu_i$, the model for the $2\times J$ transformed response $\vd_i$ is the normal with the diagonal covariance matrix  $\mOmega_i$. 
Therefore, in this conditionally normal model, the factor component term $\mLam\vu_i$ captures the association within and between the at-risk and Poisson parts over $J$ dimensions. 
Based on this representation, the following subsection introduces the parameter expansion for efficient sampling of the factor components. 

\subsection{Parameter expansion and prior distributions}
\label{sec:prior}
For the regression parameters $\vbeta_{hj}$, we assume the conditionally conjugate priors $N(\vb_0,\mB_0)$ for $h=1,2$ and $j=1,\dots,J$. 
The standard prior distributions for the common factors and loadings would be $u_{ij}\sim N(0,1)$ and $\lambda_{hj}\sim N(0,1)$. 
Under this prior specification, however, the mixing of an MCMC algorithm tends to be very slow. 

The augmented model above is expanded for efficient posterior sampling, borrowing the idea of  \cite{GD09}. 
Specifically, we introduce the working parameters $\vlam_{hj}^*=(\lam^*_{hj1},\dots,\lam^*_{hjK})'$ and $\vu_i^*=(u^*_{i1},\dots,u^*_{iK})'$. 
The MCMC algorithm samples the working parameters from their posterior distributions. 
The likelihood contribution of the $i$th individual in the expanded model is obtained by simply replacing $\vu_{i}$ and $\vlam_{hj}$ in \eqref{eqn:Li} with $\vu_{i}^*$ and $\vlam_{hj}^*$, respectively. 
The prior distributions for the working parameters are given by  $\lambda_{hjk}^*\sim N(0,1),\ h=1,2, \ j=1,\dots,J, \ k=1,\dots,K$, and 
$\vu_i^*\sim N(\zero, \mPhi), \ i=1,\dots,N$ where
$\mPhi=\diag(\phi_1,\dots,\phi_K)$. 
Further, it is assumed 
$\phi_k\sim IG(a_k,b_k),\ k=1,\dots,K$. 

Our prior specification differs slightly from that in \cite{GD09}. 
To correct for the invariance of the factor loadings due to rotation and sign-switching,  \cite{GD09} employed the positive lower triangular (PLT) constraint where the diagonal elements of the factor loading matrix are strictly positive, and the upper triangle elements are fixed to zero a-priori. 
In our model, it would have been $\lambda_{1jk}^*=0,\ k=\min(j,K)+1,\dots,K$ and $\lambda_{2jk}^*=0,\ k=\min(J+j,K)+1,\dots,K$. 
However, PLT only partially solves the identification issues. 
For example, the identifiability is lost when the loading for the first variable is close to zero. 
In this case, reordering the variables is required. 
See \cite{Papa} and references therein for the recent development in the approaches to achieving identifiability of the factor model and their limitations. 

Therefore, this paper employs the parameter expansion without constraining the factor loading matrix. 
The MCMC draws of the unidentified parameters are post-processed to produce the posterior draws of the identified parameters.
See Section~\ref{sec:post_process}.

\subsection{MCMC algorithm}\label{sec:mcmc}
The parameters and latent variables are sampled using the Gibbs sampler described in the following. 
In this section, $\eta_{hij}$ is expressed in terms of  the working parameters $\eta_{hij} = \vx_{ij}'\vbeta_{hj} + \vu_i^{*'}\vlam_{hj}^*$. 

The joint distribution of the parameters and latent variables under the expanded model is proportional to
\begin{equation}
\begin{split}
&\prod_{j=1}^J \prod_{i=1}^N\left[
    \exp\left\{-\frac{\omega_{1ij}}{2}(\vx_{i}'\vbeta_{1j}+\vu_i^{*'}\vlam^*_{1j})^2
    +\kappa_{1ij}(\vx_{i}'\vbeta_{1j}+\vu_i^{*'}\vlam^*_{1j})\right\}p(\omega_{1ij})\right.\\
&\quad\times\left.\left(\exp\left\{-\frac{\omega_{2ij}}{2}(\vx_{i}'\vbeta_{2j}+\vu_i^{*'}\vlam^*_{2j}-\log r)^2+\kappa_{2ij}(\vx_{i}'\vbeta_{2j}+\vu_i^{*'}\vlam^*_{2j}-\log r)\right\}p(\omega_{2ij})\right)^{I(z_{ij}=1)I(y_{ij}=g,y_{ij}^*\in[\kappa_g,\kappa_{g+1}))} \right]\\
&\quad\times \left[\prod_{i=1}^Np(\vu_i^*|\mPhi)\right] \left[\prod_{h=1}^2\prod_{j=1}^J\prod_{k=1}^Kp(\vlam^*_{hjk})\right]
\left[\prod_{h=1}^2\prod_{j=1}^Jp(\vbeta_{hj})\right]p(\mPhi)
\end{split}    
\end{equation}
where  $I(\cdot)$ is the indicator function, $p(\vlam_{hjk}^*)$, $p(\vbeta_{hj})$ and $p(\Phi)$ denote the prior densities. 
Then, the Gibbs sampler alternately samples $\{y^*_{ij}\}$, $\{\vbeta_{hj}\}$, $\{z_{ij}\}$, $\{\omega_{ij}\}$, $\{\vlam_{hj}^*\}$, $\{\vu_i^*\}$ and $\{\phi_k\}$ from their respective full conditional distributions.

\begin{enumerate}
\item
Sampling $y_{ij}^*,\ i=1,\dots,N, \ j=1,\dots,J$: From  \eqref{eqn:grouping} and conditionally on $z_{ij}=1$, after integrating $\omega_{2ij}$, the full conditional distribution of $y_{ij}^*$ is proportional to
\[
p(y_{ij}^*|z_{ij}=1,\text{Rest})\propto \frac{\Gamma(y_{ij}^*+r)}{\Gamma(r)y_{ij}^*!}\frac{(e^\psi_{ij})^{y_{ij}^*}}{(1+e^{\psi_{ij}})^{y_{ij}^*+r}}I(y_{ij}=g,y_{ij}^*\in[\kappa_g,\kappa_{g+1})),
\]
where $\psi_{ij}=\eta_{2ij}-\log r$. 
This full conditional distribution is the negative 
binomial distribution truncated on the interval $[\kappa_g,\kappa_{g+1})$. 

\item
The sampling steps of $z_{ij}$,  $\vbeta_{hj}$, $\vlam_{hj}^*$ and $\omega_{hij}$  are similar to those provided in \citet{Neelon19}. 
\begin{itemize}
    \item
    Sampling $z_{ij},\ i=1,\dots,N, \ j=1,\dots,J$: The full conditional distribution of $z_{ij}$ is given by
    \[
        \Pr(z_{ij}=1|y_{ij}^*=0, \text{Rest})=\frac{\pi_{ij}v_{ij}^r}{1-\pi_{ij}(1-v_{ij}^r)},
    \]
    where $v_{ij}=1/(1+e^{\psi_{ij}})$. 
    \item
    Sampling $\omega_{hij},\ h=1,2,\ i=1,\dots,N,\  j=1,\dots,J$: $\omega_{1ij}$ is sampled from $PG(1,\eta_{itj})$. 
    Similarly, for $i$ and $j$ such that $z_{ij}=1$, $\omega_{2ij}$ is sampled from $PG(r+y_{ij}^*, \eta_{2ij}-\log r)$
    \item 
    Sampling $\vbeta_{1j}$ and $\vlam_{1j}^*, \ j=1,\dots,J$: 
    We sample $\vbeta_{1j}$ and $\vlam_{1j}^*$ in one block. 
    The full conditional distribution of $(\vbeta_{1j}',\vlam_{1j}^{*'})'$ is given by $N(\vb_{1j},\mB_{1j})$ where 
    \[
        \mB_{1j} = \left[\sum_{i=1}^N\omega_{1ij}\tilde{\vx}_{ij}\tilde{\vx}_{ij}'+\tilde{\mB}_{0}^{-1}\right]^{-1}, \quad
        \vb_{1j} =\mB_{1j}\left[\sum_{i=1}^N\tilde{\vx}_{ij}\left(z_{ij}-\frac{1}{2}\right)+\tilde{\mB}_{0}^{-1}\tilde{\vb}_{0}\right],
    \]
    where  $\tilde{\vx}_{ij}=(\vx_{ij}',\vu_{i}^{*'})'$, $\tilde{\mB}_0$ is the block diagonal matrix with $\mB_0$ and $\mI_{\ell}$  on the diagonal blocks and $\tilde{\vb}_0=(\vb_0',\zero_{K}')'$.  
    
    \item
    Sampling $\vbeta_{2j}$ and $\vlam_{2j}^*, \ j=1,\dots,J$: Similarly, $\vbeta_{2j}$ and $\vlam_{2j}^*$ are sampled in one block. 
    The full conditional distribution of $(\vbeta_{2j}',\vlam_{2j}^{*'})'$ is given by $N(\vb_2,\mB_2)$ where
    \[
        \mB_{2j}=\left[\sum_{i:z_{ij}=1}\omega_{2ij}\tilde{\vx}_{ij}\tilde{\vx}_{ij}+\tilde{\mB}_{0}^{-1}\right]^{-1},\quad
        \vb_{2j}=\mB_{2j}\left[\sum_{i:z_{ij}=1}\tilde{\vx}_{ij}\left(\frac{y_{ij}^*-r}{2}+\omega_{2ij}\log r\right)+\tilde{\mB}_{0}^{-1}\tilde{\vb}_{0}\right], 
    \]
        
\end{itemize}

\item
Sampling $\vu_i^*,\ i=1,\dots,N$: 
The full conditional distribution of $\vu_i^*$ is $N(\vm_i,\mV_i)$ where
\[
\begin{split}
\mV_i&=\left[\sum_{j=1}^J\omega_{1ij}\vlam^*_{1j}\vlam^{*'}_{1j}+\sum_{j:z_{ij}=1}\omega_{2ij}\vlam^*_{2j}\vlam^{*'}_{2j}+\mPhi^{-1}\right]^{-1},\\
\vm_i&=\mV_i\left[\sum_{j=1}^J(\kappa_{1ij}-\omega_{1ij}\vx_i'\vbeta_{1j})\vlam_{1j}^*+\sum_{j:z_{ij}=1}(\kappa_{2ij}-\omega_{2ij}(\vx_i'\vbeta_{2j}-\log r))\vlam_{2j}^*\right]
\end{split}
\]

\item
Sampling $\phi_k,\ k=1,\dots,K$: The full conditional distribution of $\phi_k$ is given by $IG(a_k+N/2, b_k+\sum_{i=1}^N u_{ik}^{*2}/2)$
\end{enumerate}

\subsection{Post-processing}\label{sec:post_process}
The MCMC draws of the factor components are processed in the following two steps. 
First, the sampled working parameters are not identified in terms of scale (Section~\ref{sec:prior}). 
The original parameters are recovered through 
\begin{equation}
    \lam_{hjk}=
\lam_{hjk}^*\phi_k^{1/2},\quad u_{ik}=
u_{ik}^*\phi_k^{-1/2},\quad
j=1,\dots,J, \quad k=1,\dots,K. 
\end{equation}
Then, these parameters are still subject to the rotational and sign-switching invariance. 
We apply the post-processing algorithm of \cite{Papa} to the MCMC draws of $\lambda_{hjk}$. 
The algorithm first applies the varimax rotation to each MCMC draw to solve the rotational invariance, then to solve the sign-switching invariance, it applies the signed permutations to the MCMC output until the transformed loadings are sufficiently close to some reference value. 
Their algorithm is provided in the R package \texttt{factor.switching}. 
See \cite{Papa} for details.

\section{Simulation study}\label{sec:sim}
Here, the performance of the proposed model is investigated using the simulated data. 
We set $N=1000$, $J=10$, $K=1$ and $P=2$. 
The regression coefficients are given by $\vbeta_{1j}^\text{true}=(0.5,0.5)$, $\vbeta_{2j}^\text{true}=(-0.5,-1)$ for $j=1,\dots,J$. 
The covariate vector is $\vx_{ij}=(1,x_i)'$ for $i=1,\dots,N$, $j=1,\dots,J$, and $x_i\sim N(0,1)$. 
For the factor loadings, $\vlam_1^\text{true}=(0.89,0,0.25,0,0.8,0,0.5,0,0,0)'$ and $\vlam_2^\text{true}=(0,0,0.85,0.8,0,0.75,0.75,0,0.8,0.8)'$. 
We consider the following two settings for the grouping mechanisms. 
In Setting~1, it is set $\{0\}$, $\{1\}$, $\{2\}$, $[3,5]$, $[6,10]$, $[11,50]$, $[51,\infty)$, which is the same as  the NLSY79 data in Section~\ref{sec:real}. 
Setting~2 considers the finer grouping mechanism  such that the grouped data contain more information: $\{0\}$,  $\{1\}$, $\{2\},\dots,\{10\}$, $[11,15]$, $[16,20]$, $[21,25]$, $[26,30]$, $[31,40]$, $[41,50]$, $[51,\infty)$. 
The data are replicated $R=100$ times. 
The overall proportion of structural zeros is approximately $0.6$.

The proposed factor ZIP model for grouped data (GFZIP) is compared with the following three models. 
Firstly, the ZIP model for grouped data (GZIP) is considered. 
Since this model does not include factors that provide links among structural zeros and grouped counts, it is essentially a univariate model and thus is estimated separately for each $j$. 
Secondly, ZINB \citep{McG17} for grouped data is also considered. 
Finally, to assess the effect of the loss of information due to the grouping mechanism, the factor ZIP (FZIP) model for the ungrouped count data is considered and fitted to the underlying count data without the grouping mechanism. 

For all models, we assume $\vbeta_{hj}\sim N(\zero, 100\mI)$ for $h=1,2$ and $j=1,\dots,J$. 
For each model, the MCMC algorithm is run for 22,000 iterations, with the initial 2,000 draws discarded as the burn-in period. 
The parameter estimation is based on the remaining 20,000 MCMC draws. 

The performance of the models is assessed based on the bias $\text{Bias}(\beta_{hjp})=\frac{1}{R}\sum_{r=1}^R\left(\hat{\beta}^{(r)}_{hjp}-\beta_{hjp}^\text{true}\right)$ 
and root mean squared errors (RMSE) $\text{RMSE}(\beta_{hjp})=\sqrt{\frac{1}{R}\sum_{r=1}^R\left(\hat{\beta}^{(r)}_{hjp}-\beta_{hjp}^\text{true}\right)^2}$ for $h=1,2$, $j=1,\dots,J$ and $p=1,\dots,P$, where $\hat{\beta}_{hjp}^{(r)}$ is the posterior mean from $r$th replication of the data. 
For the factor loadings, we compute the bias and RMSE for $\text{vec}(\mLam\mLam')$, as the post-processed signs of the loadings vary over the replications. 

We also evaluate the true positive (TPR), true negative (TNR), false positive (FPR) and false negative (FNR) rates for being at-risk conditionally on the zero response. 
The posterior probability of $i$th individual being at-risk in $j$th dimension given the response $y_{ij}=0$ is denoted by $\Pr(z_{ij}=1|y_{ij}=0)$. 
It is estimated by $\hat{\pi}_{ij}=\frac{1}{M}\sum_{m=1}^M z_{ij}^{m}$ for $i$ such that $y_{ij}=0$ based on the $M$ draws of the MCMC algorithm. 
The individual $i$ is deemed to be at-risk in $j$th dimension if $\hat{\pi}_{ij}>0.5$. 
Then, the TPR, TNR, FPR, and FNR are calculated as
\[
\begin{split}
\text{TPR}_j = \frac{\sum_{i=1}^NI(\hat{\pi}_{ij}>0.5, y_{ij}=0, z^{true}_{ij}=1)}{\sum_{i=1}^N I(y_{ij}=0, z^{true}_{ij}=1)},\quad 
\text{TNR}_j = \frac{\sum_{i=1}^NI(\hat{\pi}_{ij}\leq0.5, y_{ij}=0, z^{true}_{ij}=0)}{\sum_{i=1}^N I(y_{ij}=0, z^{true}_{ij}=0)}\\
\text{FPR}_j = \frac{\sum_{i=1}^NI(\hat{\pi}_{ij}>0.5, y_{ij}=0, z^{true}_{ij}=0)}{\sum_{i=1}^N I(y_{ij}=0, z^{true}_{ij}=0)},\quad 
\text{FNR}_j = \frac{\sum_{i=1}^NI(\hat{\pi}_{ij}\leq0.5, y_{ij}=0, z^{true}_{ij}=1)}{\sum_{i=1}^N I(y_{ij}=0, z^{true}_{ij}=1)},
\end{split}
\]
for $j=1,\dots,J$, 
where  $z^{true}_{ij}$ denotes the true value of the latent at-risk indicator $z_{ij}$. 

As in the real data application on the youths' involvement in illegal activities in Section~\ref{sec:real}, the proportion of at-risk individuals among those whose responses are zero would be a quantity of interest. 
The proportion of interest is defined by
\begin{equation}\label{eqn:Rj}
\hat{R}_j\equiv \frac{\sum_{i=1}^N I(\hat{\pi}_{ij}>0.5, y_{ij}=0)}{\sum_{i=1}^N I(y_{ij}=0)},\quad j=1,\dots,J. 
\end{equation}

Tables~\ref{tab:sim_beta}  presents the biases and RMSEs for the coefficients $\vbeta_{1j}$ and $\vbeta_{2j}$ from 100 replications of the data averaged over $J$ dimensions. 
Since FZIP, which knows the true underlying counts before grouping, is not affected by the grouping mechanism, it produces identical results under both simulation settings. 
Therefore, the cells for FZIP in Setting~2 are left blank.

When comparing the proposed GFZIP, GZIP, and GZINB, ignoring the factor structure among the at-risk probabilities and Poisson parts leads to larger bias and RMSE. 
As expected, the GFZIP performed the best among the three models. 
GZINB resulted in large RSME for the at-risk coefficients  $\vbeta_{1jp}$, especially in the case of Setting~1. 
This is due to the numerical instability from the coarse grouped data. 
Compared to FZIP, GFZIP resulted in increased bias and RMSE for the Poisson coefficients $\vbeta_{1j}$ due to the loss of information through the grouping mechanism. 
It is also seen that the performance of GFZIP regarding the Poisson coefficients improves as the number of groups increases from Setting~1 to Setting~2, where the grouped data contain more information. 
This is also the case for GZIP and GZINB, and this phenomenon was also observed in \cite{McG17}. 

\begin{figure}[H]
\centering
\includegraphics[width=0.95\textwidth]{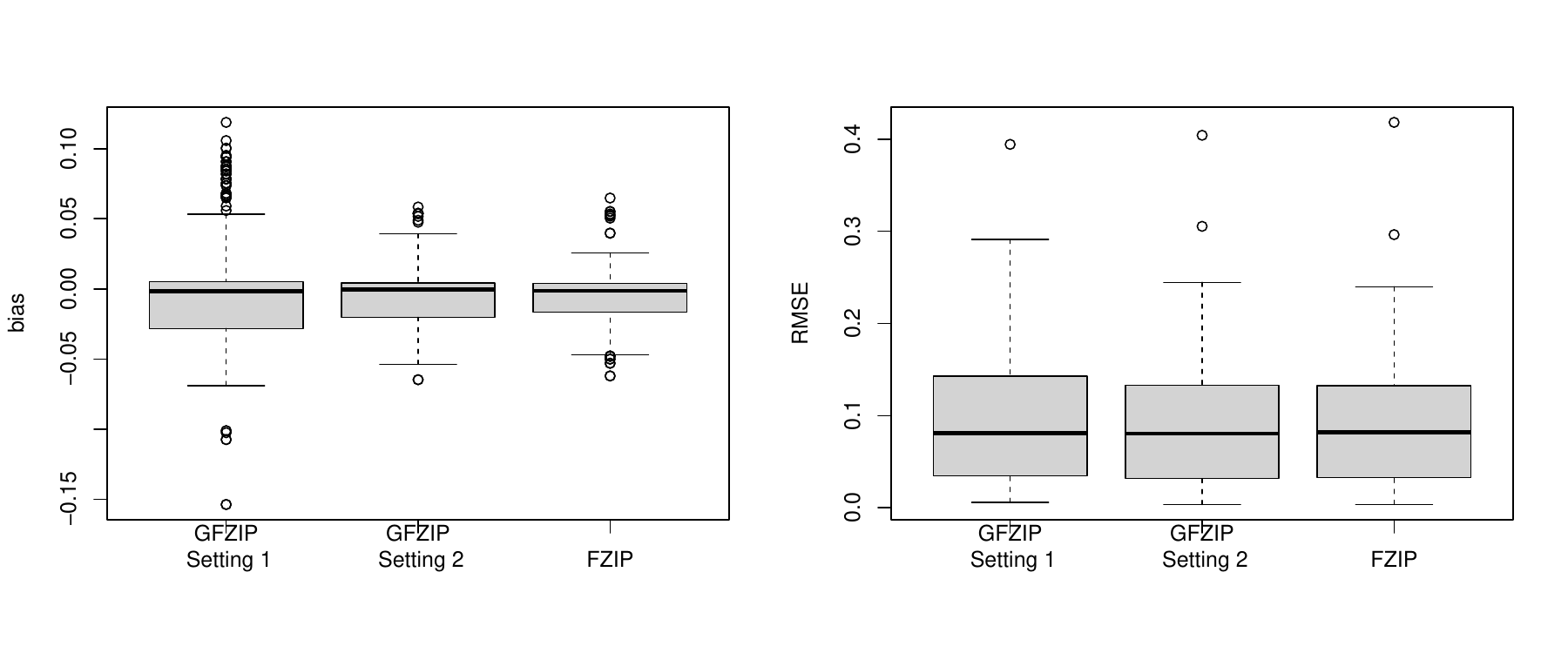}
\caption{Boxplots of bias and RMSE for $\text{vec}(\mLam\mLam')$}
\label{fig:LL}
\end{figure}

Figure~\ref{fig:LL} presents the boxplots of the bias and RMSE for $\text{vec}(\mLam\mLam')$ under GFZIP and FZIP. 
The bias under GFZIP is larger than that under FZIP in both settings due to grouping. 
It is also seen that the bias under GFZIP decreases as the finer grouping mechanism is used in Setting~2. 
A similar pattern is observed for the RMSE.

\begin{figure}[H]
    \centering
    \includegraphics[width=0.95\textwidth]{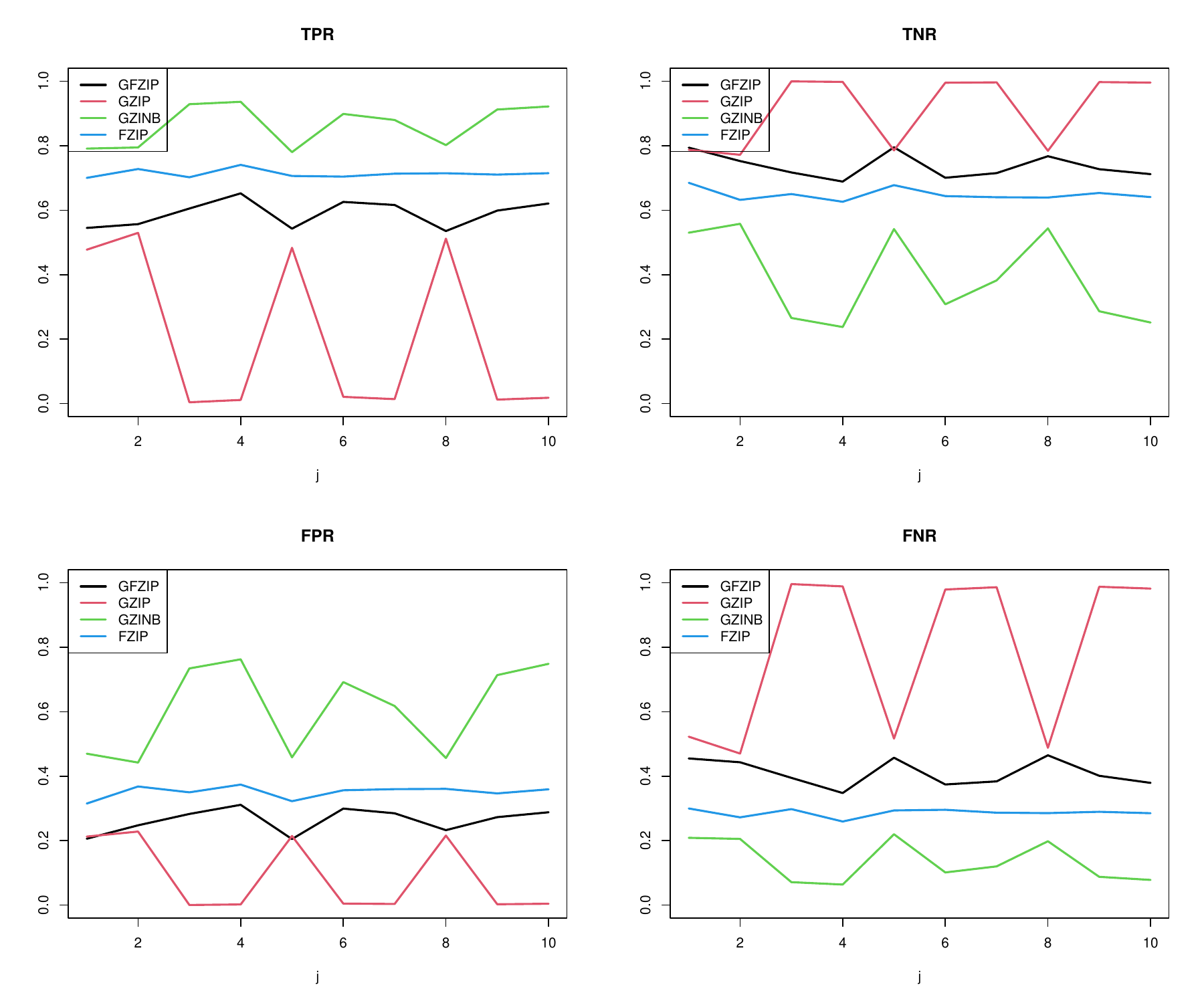}
    \caption{True positive (TPR), true negative (TNR), false positive (FPR) and false negative rates (FNR) for GFZIP, GZIP, GZINB, and FZIP in Setting~1}
    \label{fig:pp1}
\end{figure}

\begin{figure}[H]
    \centering
    \includegraphics[width=0.95\textwidth]{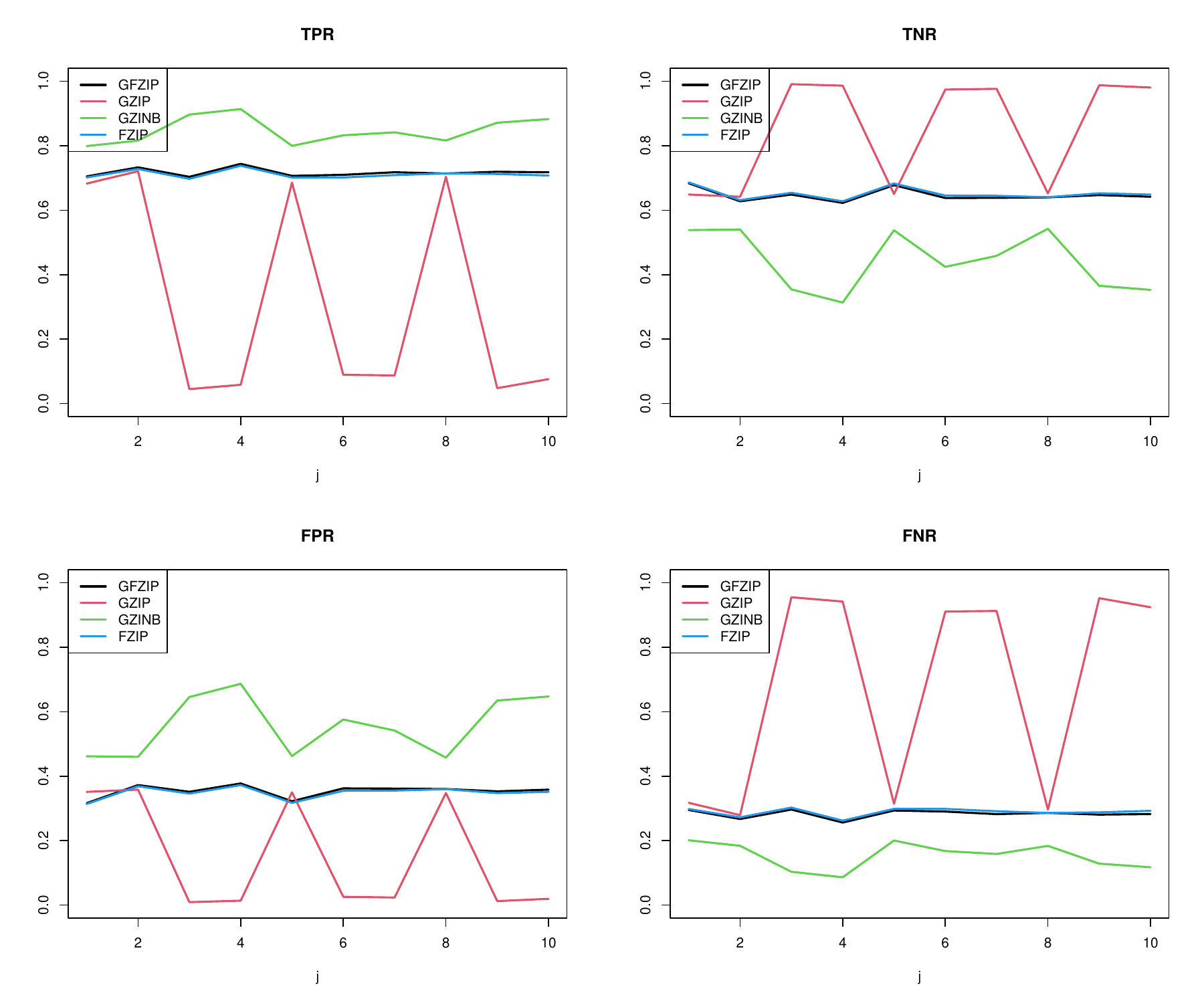}
    \caption{True positive (TPR), true negative (TNR), false positive (FPR) and false negative rates (FNR) for GFZIP, GZIP, GZINB, and FZIP in Setting~2}
    \label{fig:pp2}
\end{figure}

Figures~\ref{fig:pp1} and \ref{fig:pp2} present TPR, TNR, FPR, and FNR averaged over 100 replications under GFZIP, GZIP, GZINB and FZIP in Settings~1 and 2, respectively. 
Firstly, we observe that the results under GFZIP and FZIP become almost identical in Setting~2, while there are some discrepancies in Setting~1. 
In both settings, GZIP resulted in TPR and FPR for some dimensions being close to zero. 
On the contrary, TNR and FNR in those dimensions are close to one. 

\begin{figure}[H]
    \centering
    \includegraphics[width=0.45\textwidth]{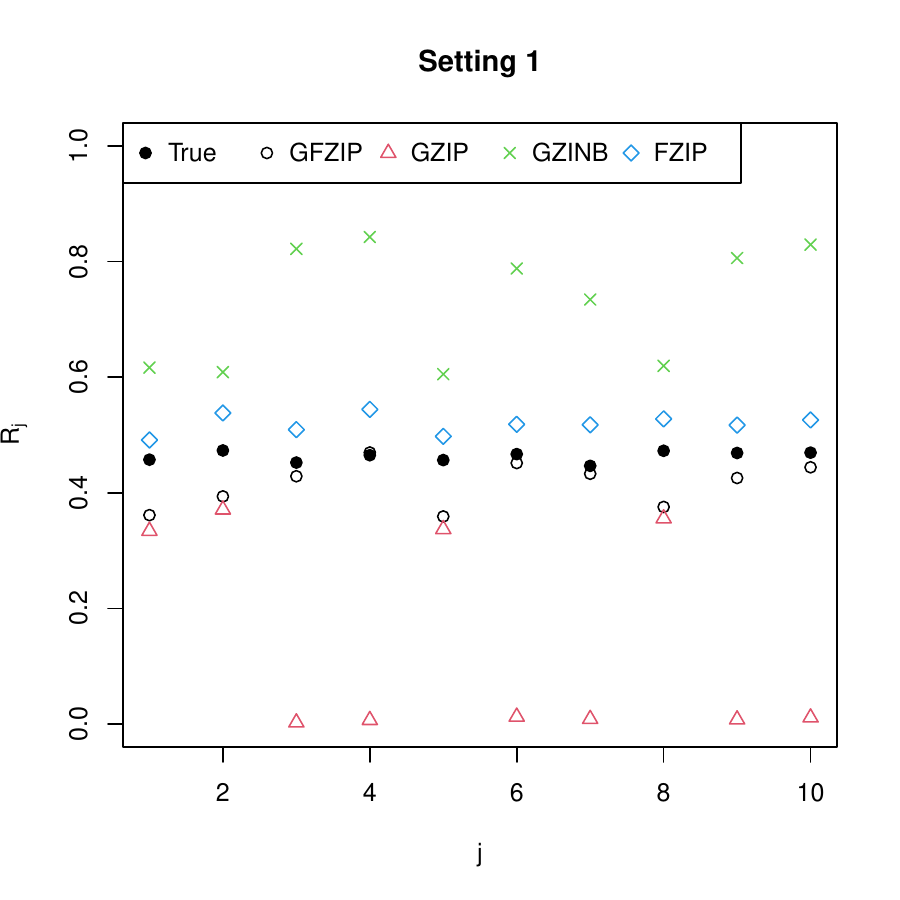}
    \includegraphics[width=0.45\textwidth]{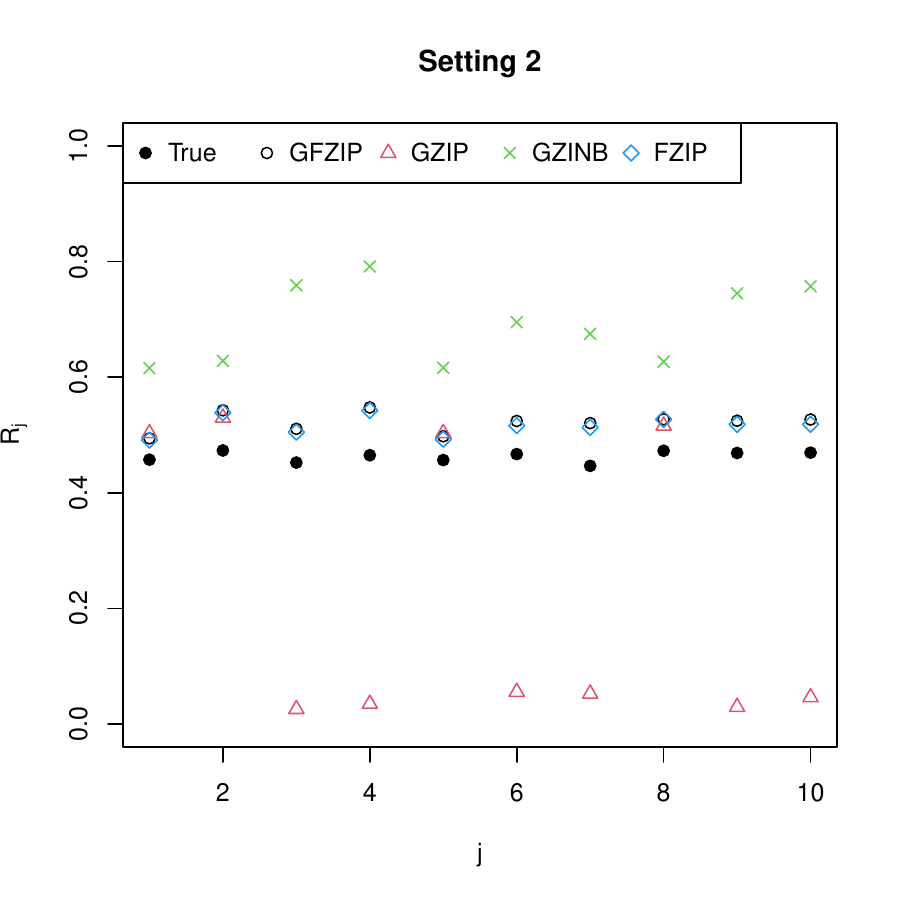}
    \caption{At-risk proportions among those with $y_{ij}=0$, $\hat{R}_j$}
    \label{fig:pz}
\end{figure}

Figure~\ref{fig:pz} presents the estimated proportions of at-risk individuals given $y_{ij}=0$, $\hat{R}_j$, averaged over 100 replications. 
The GFZIP and FZIP models seem to work well, with the estimates being close to the truth. 
Their results become almost identical in Setting~2, similar to Figures~\ref{fig:pp1} and \ref{fig:pp2}. 
The figure also shows that GZINB overestimates $R_j$ in both Settings. 
The results for GZIP are similar to those in the previous figures. 
In some dimensions, the estimates of $R_j$ under GZIP are close to zero, implying the false negative rates are close to one. 
In the other dimensions, the estimates under GZIP are similar to those of GFZIP. 
The behaviour in GZIP results from ignoring the association between the dimensions. 
In the real data analysis in Section~\ref{sec:real}, where it would be natural to consider the association between the youths' illegal activities, we observe a similar result under GZIP.

\begin{table}[H]
    \centering
    \caption{Bias and RMSE for the at-risk coefficients $\beta_{1jp}$ from 100 replications averaged over $J=10$ dimensions. The results for FZIP, which are not affected by the grouping mechanism, are the same for both settings.}
    \label{tab:sim_beta}
    \begin{tabular}{ccrrrrrrrr}\toprule
    &&\multicolumn{4}{c}{Bias} & \multicolumn{4}{c}{RMSE}\\
    \cmidrule(lr){3-6}\cmidrule(lr){7-10}
    Parameter & Setting & GFZIP & GZIP & GZINB & FZIP& GFZIP & GZIP & GZINB & FZIP\\\hline
$\beta_{1.1}$& 1& -0.093 & -0.618 &  1.549 &  0.110 & 0.290 & 0.667 &  2.320 & 0.303\\
$\beta_{1.2}$&  & -0.043 & -0.329 &  0.389 &  0.067 & 0.205 & 0.382 &  0.797 & 0.218\\
$\beta_{1.1}$& 2&  0.115 & -0.428 &  1.044 & ------ & 0.302 & 0.565 &  1.680 & ------\\
$\beta_{1.2}$&  &  0.071 & -0.210 &  0.330 & ------ & 0.217 & 0.315 &  0.633 & ------\\
\hline 
$\beta_{2j1}$& 1&  0.087 &  0.533 &  0.084 & -0.023 & 0.153 & 0.555 &  0.194 & 0.111\\
$\beta_{2j2}$&  & -0.036 & -0.178 & -0.029 & -0.006 & 0.076 & 0.163 &  0.098 & 0.062\\
$\beta_{2j1}$& 2& -0.025 &  0.394 &  0.048 & ------ & 0.111 & 0.449 &  0.188 & ------\\
$\beta_{2j2}$&  & -0.009 & -0.076 & -0.007 & ------ & 0.063 & 0.125 &  0.083 & ------\\
\bottomrule
\end{tabular}
\end{table}

\section{Analysis of illegal activities of youth}\label{sec:real}
\label{sec:real}
\subsection{Data and setting}
We consider the number of times youths were involved in the nineteen illegal activities  ($J=19$) obtained from the 1980 round of the National Longitudinal Study of Youth 1979 (NLSY79) data. 
In NLSY79, the questionnaire was designed so that the respondents answered at an exact frequency or interval of frequencies of each illegal activity in the year prior to the interview. 
Then, the answers are published as the grouped count data. 
Although this is old data, it provides valuable information on the problematic behaviour of youths, of which statistical analyses are still relevant today.

The choices are `never' ($g=0$: $\{0\}$), `once'  ($g=1$: $\{1\}$), `twice' ($g=2$: $\{2\}$), `between three and five times' ($g=3$: $[3,5]$), `between six and ten times' ($g=4$: $[6,10]$), `between eleven and fifty times' ($g=5$: $[11,50]$) and `over fifty times' ($g=6$: $[51,\infty]$). 
Table~\ref{tab:nlsy_act} describes the nineteen activities considered in this analysis and associated labels used in the following figures and tables.

Figure~\ref{fig:nlsy_hist} presents the histograms of the times $2865$ youths were involved in the 19 illegal activities in the previous year. 
The numbers in the panels indicate the fractions of zeros. 
A substantially large proportion of youths were not involved in each activity, exhibiting many zeros. 
For example, the observed proportions of zeros for the activities with highly criminal nature, such as \texttt{sell\_marijuana}, \texttt{sell\_hard\_drugs} and \texttt{break\_in}, are above $0.9$, and are particularly high. 
The proportion of zeros for \texttt{alcohol} is $0.39$.
It is much lower than those for other activities as it is more common for youths, though this value may be relatively high in the context of zero-inflated count data.

\begin{figure}[H]
    \centering
    \includegraphics[width=0.32\textwidth]{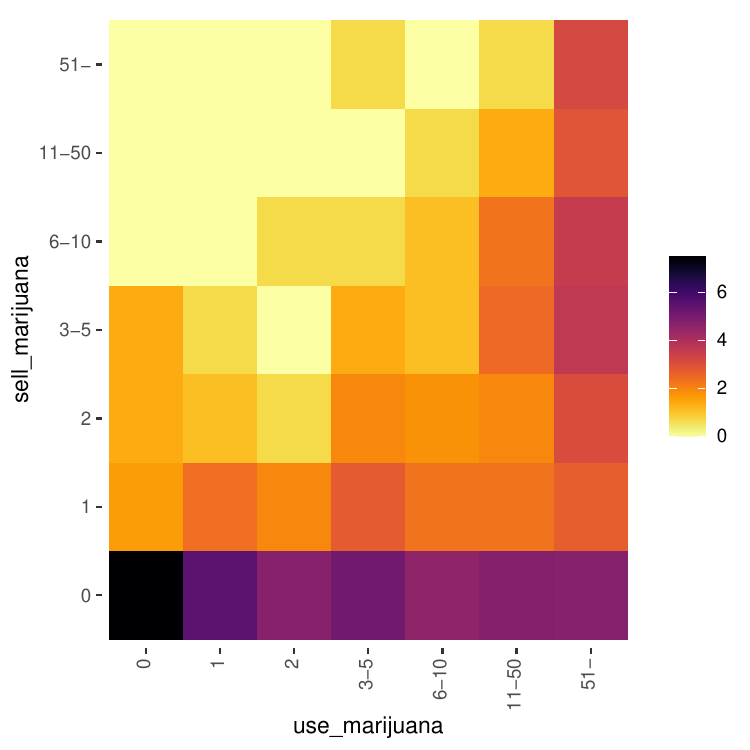}
    \includegraphics[width=0.32\textwidth]{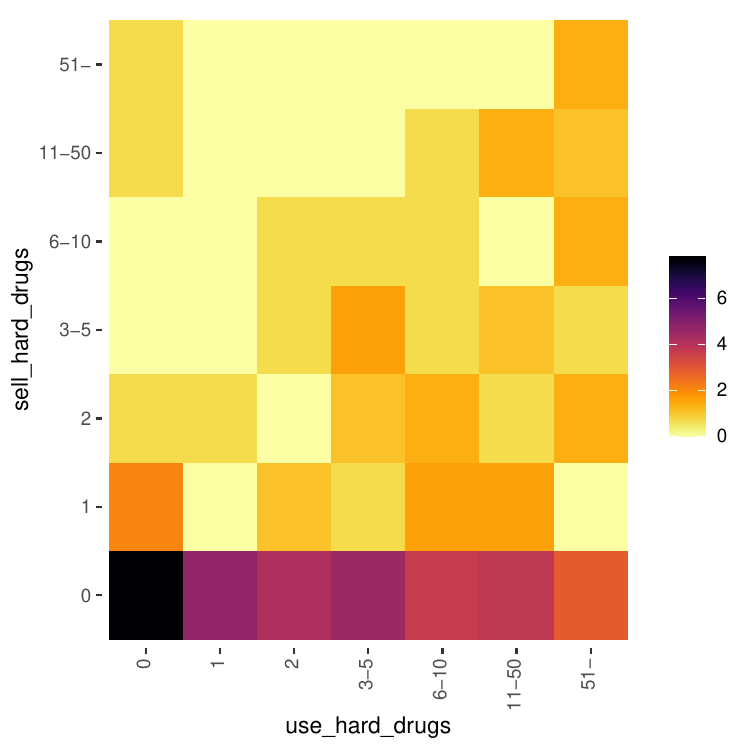}
    \includegraphics[width=0.32\textwidth]{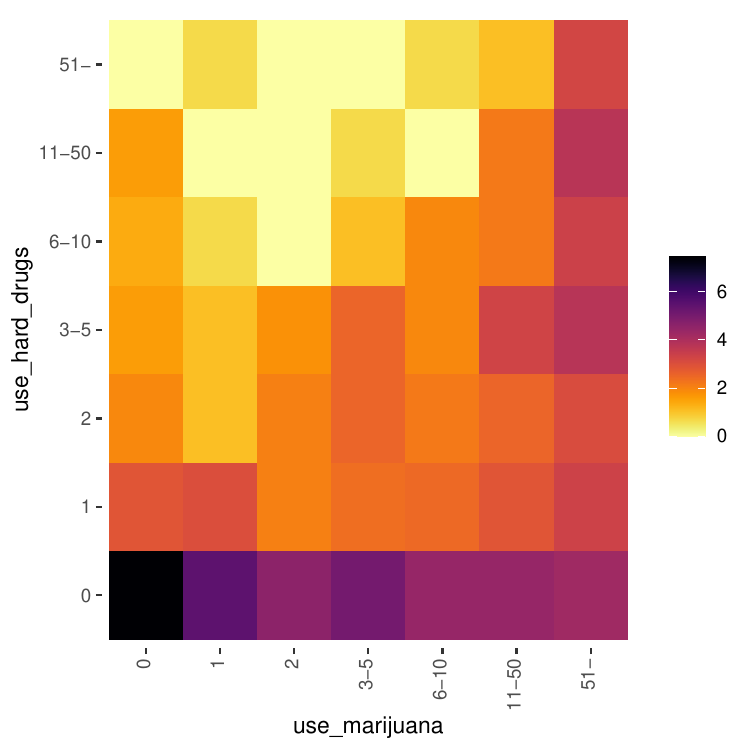}
    \includegraphics[width=0.32\textwidth]{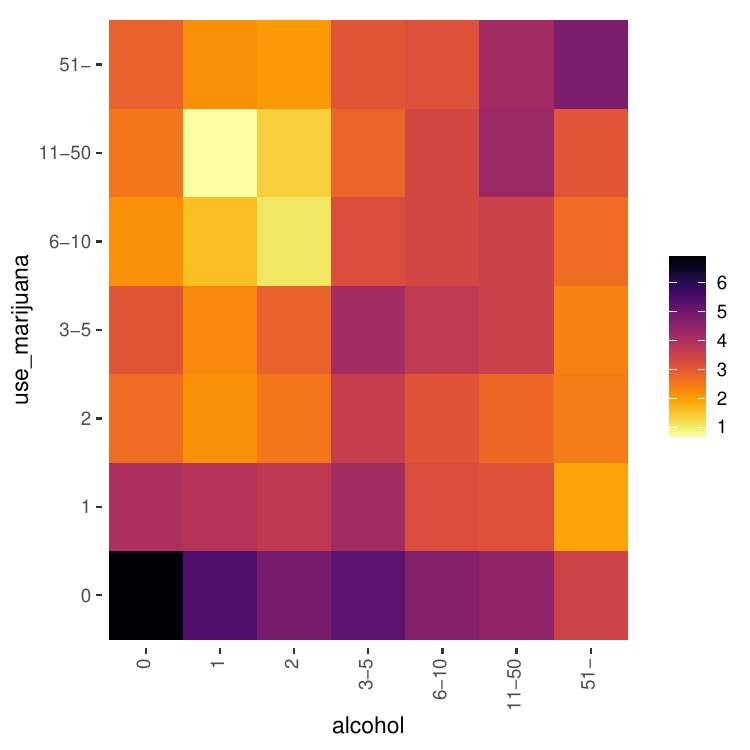}
    \includegraphics[width=0.32\textwidth]{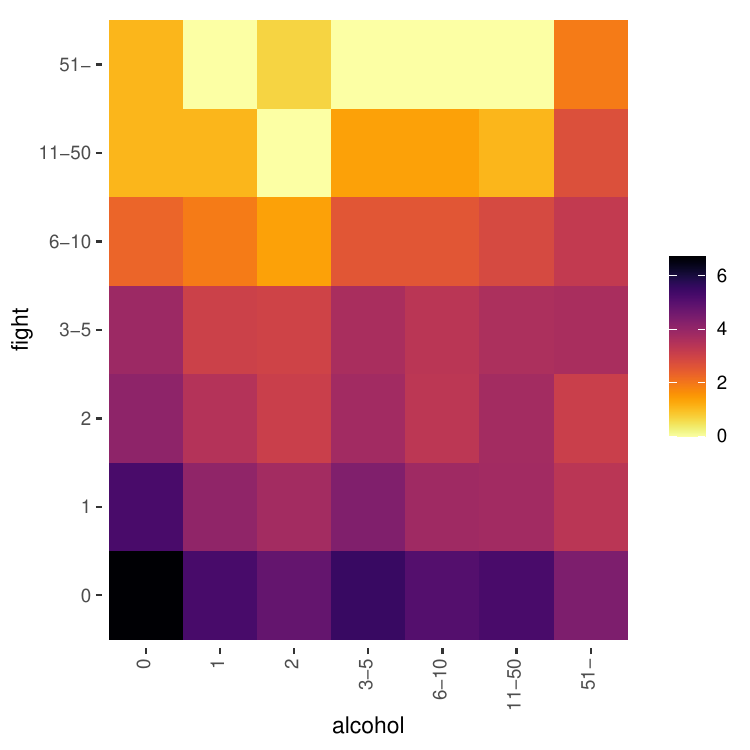}
    \includegraphics[width=0.32\textwidth]{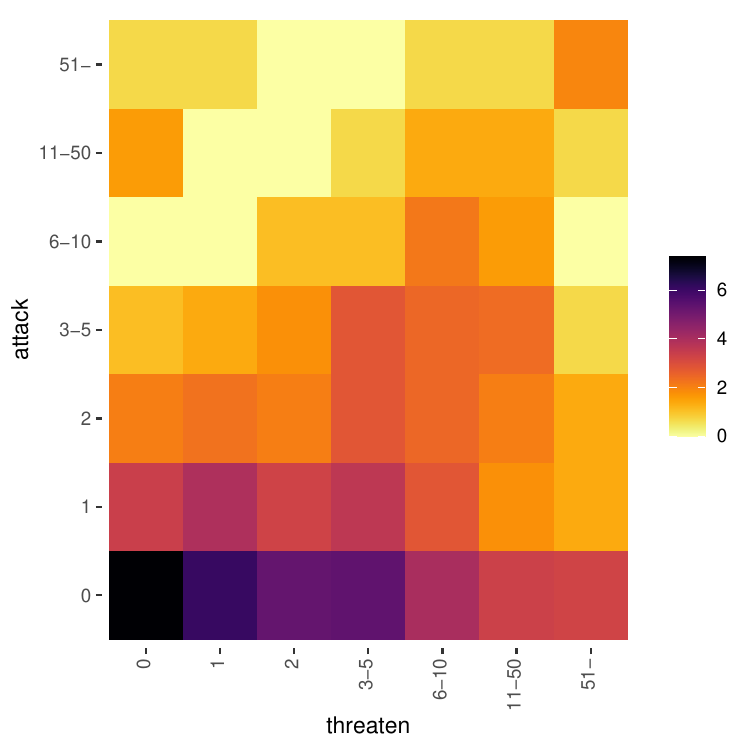}
    \caption{Heatmaps of log frequencies for the arbitrarily selected pairs of activities}
    \label{fig:nlsy_heat}
\end{figure}

The histograms only reveal the distribution of involvement in each activity separately and the extent of the zero inflation. 
However, we are also interested in the association among the activities because it would be natural to assume that involvement in one activity and its frequency may be associated with those in another activity, such as the use of alcohol and marijuana. 
Figure~\ref{fig:nlsy_heat} presents the heatmaps of log frequencies for the arbitrarily selected pairs of activities. 
The frequencies are added with one before taking the log. 
Some observations from the figure are as follows. 
The frequencies for non-involvement in neither are the highest for all pairs of activities. 
The top left and middle panels indicate that a certain fraction of youths had experience using marijuana or hard drugs while they did not sell them. 
The top left panel also shows that the youths who sold marijuana frequently used marijuana frequently, as indicated by the darker shades in the top right corner of the panel. 
A similar pattern is seen in the pair of hard drugs and marijuana in the top right panel, where most youths tended to use marijuana only, but the frequent users used both of them. 
The bottom left panel shows that frequent drinking of alcohol is associated with frequent use of marijuana, indicating they may be used together. 
Therefore, it would be more appropriate to analyse all activities jointly rather than treat each separately. 
The proposed GFZIP model can take into account these data characteristics. 

Since the information specific to each involvement in an activity is not available, only the individual characteristics are used as covariates: $\vx_{ij}=\vx_i$ for  $i=1,\dots,N$. 
The covariate information includes the constant, age, gender, race, grade, residence, poverty and mental status. 
Table~\ref{tab:x} presents the summary of the covariates. 
For the prior distributions for the coefficient vectors,  we use $\vbeta_j\sim N(\zero, 100\mI)$ for $j=1,\dots,J$. 

As in the simulation study, we compare the proposed GFZIP model with GZIP and GZINB models. 
For GFZIP, we consider the three cases for the number of factors: $K=1,2,3$. 
For each model, the MCMC algorithm is run for 60,000 iterations. 
The first 20,000 draws are discarded as burn-in period and the remaining 40,000 draws are retaind for the posterior inference. 
The models are compared based on a version of the posterior predictive loss (PPL) of \cite{GG98}, which is similar to the one considered by \cite{suga}:
\[
\text{PPL}(\M) = \frac{1}{N}\sum_{j=1}^J\sum_{g=0}^G V_{jg}^\M + \frac{1}{N+1}\sum_{j=1}^J\sum_{g=0}^G(c_{jg}-E_{jg}^\M)^2
\]
where $c_{jg}$ is the number of individuals belonging to the $g$th group for the $j$th activity, and $E_{jg}^\M$ and $V_{jg}^\M$, respectively, are the mean and variance of the posterior predictive distribution for $c_{jg}$ under model $\M$. 

\begin{table}[H]
    \centering
    \caption{Illegal activities in NLSY79}
    \label{tab:nlsy_act}
    \begin{tabular}{lll}\toprule
    Label & Description\\\hline
    \texttt{alcohol} & Drank beer, wine, or liquor without parents' permission\\
    \texttt{run\_away}& Run away from home\\
    \texttt{damage} & Purposely damaged or destroyed property  \\
    \texttt{fight} & Got into a physical fight\\
    \texttt{shoplift} & Taken something from a store without paying \\
    \texttt{steal\_ lt\_\$50} & Stolen other's belongings worth less than \$50 \\
    \texttt{steal\_ ge\_\$50} & Stolen other's belongings worth equal to or more than \$50 \\
    \texttt{extort} & Used force to get money or things from a person  \\
    \texttt{threaten} & Hit or seriously threatened to hit someone \\
    \texttt{attack} & Attacked someone with the idea of seriously hurting or killing  \\
    \texttt{use\_marijuana} & Smoked marijuana or hashish \\
    \texttt{use\_hard\_drugs} & Used any drugs or chemicals except marijuana \\
    \texttt{sell\_marijuana} & Sold marijuana or hashish \\
    \texttt{sell\_hard\_drugs} & Sold hard drugs \\
    \texttt{con} & Tried to get something by lying to a person & \\
    \texttt{vehicle} & Taken a vehicle without the owner's permission \\
    \texttt{break\_in} & Broken into a building or vehicle \\
    \texttt{sell\_stolen} & Sold or held stolen goods \\
    \texttt{gambling} & Helped in a gambling operation \\\bottomrule
    \end{tabular}
\end{table}

\begin{figure}[H]
\includegraphics[width=\textwidth]{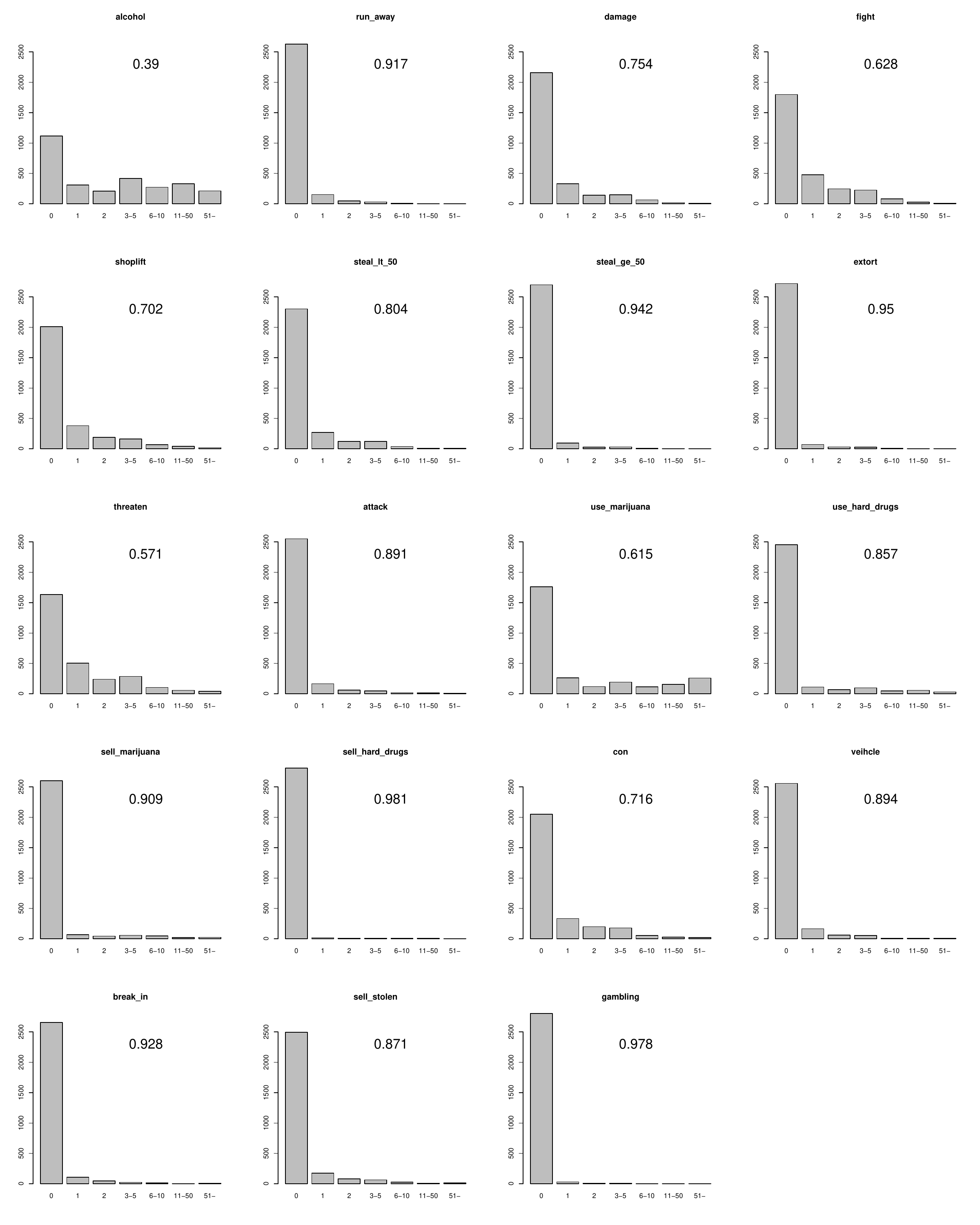}
\caption{Histograms of NLSY79 data on the illegal activities of 2865 youths. The numbers indicate the fractions of zeros.}
\label{fig:nlsy_hist}
\end{figure}

\begin{table}[H]
    \centering
    \caption{Covariates}
    \label{tab:x}
    \begin{tabular}{llrr}\toprule
       Label & Description & Mean & s.d. \\\hline
       \texttt{age}& Age of respondent & 16.11 &  0.777\\
       \texttt{male}& Dummary variable for male & 0.505 &  0.250\\
       \texttt{black} & Dummy variable for the respondent's race (black) & 0.251 & 0.434 \\
       \texttt{hisp} & Dummy variable for respondent's race (Hispanic) & 0.175 & 0.380 \\
       \texttt{grade} & Highest grade achieved & 9.555 & 1.029 \\
       \texttt{self} & Log score of self-esteem & 3.059& 0.188 \\
       \texttt{urban} & Dummy variable for the respondent living in an urban area & 0.761 & 0.428\\
       \texttt{pov} & Dummy variable for the respondent in poverty  & 0.218 & 0.413\\\bottomrule
    \end{tabular}
\end{table}

\subsection{Results}
First, we compare the posterior PPL presented in Table~\ref{tab:ppl}. 
It is shown that GFZIP with one factor resulted in the smallest PPL followed by GZIP. 
The proposed GFZIP model, which accounts for the association among the decisions on involvement with the activities and frequencies of involvements, is more appropriate than GZIP, which treats each activity separately. 
The PPL increases as the number of factors increases. 
This is a natural result, as the information in our dataset is severely limited due to the coarse grouping mechanism. 
The GZINB resulted in the largest PPL. 
This would be because the GZINB suffers from computational instability when the groups are coarsely defined, as observed in the simulation study.

\begin{figure}[H]
    \centering
    \includegraphics[width=\textwidth]{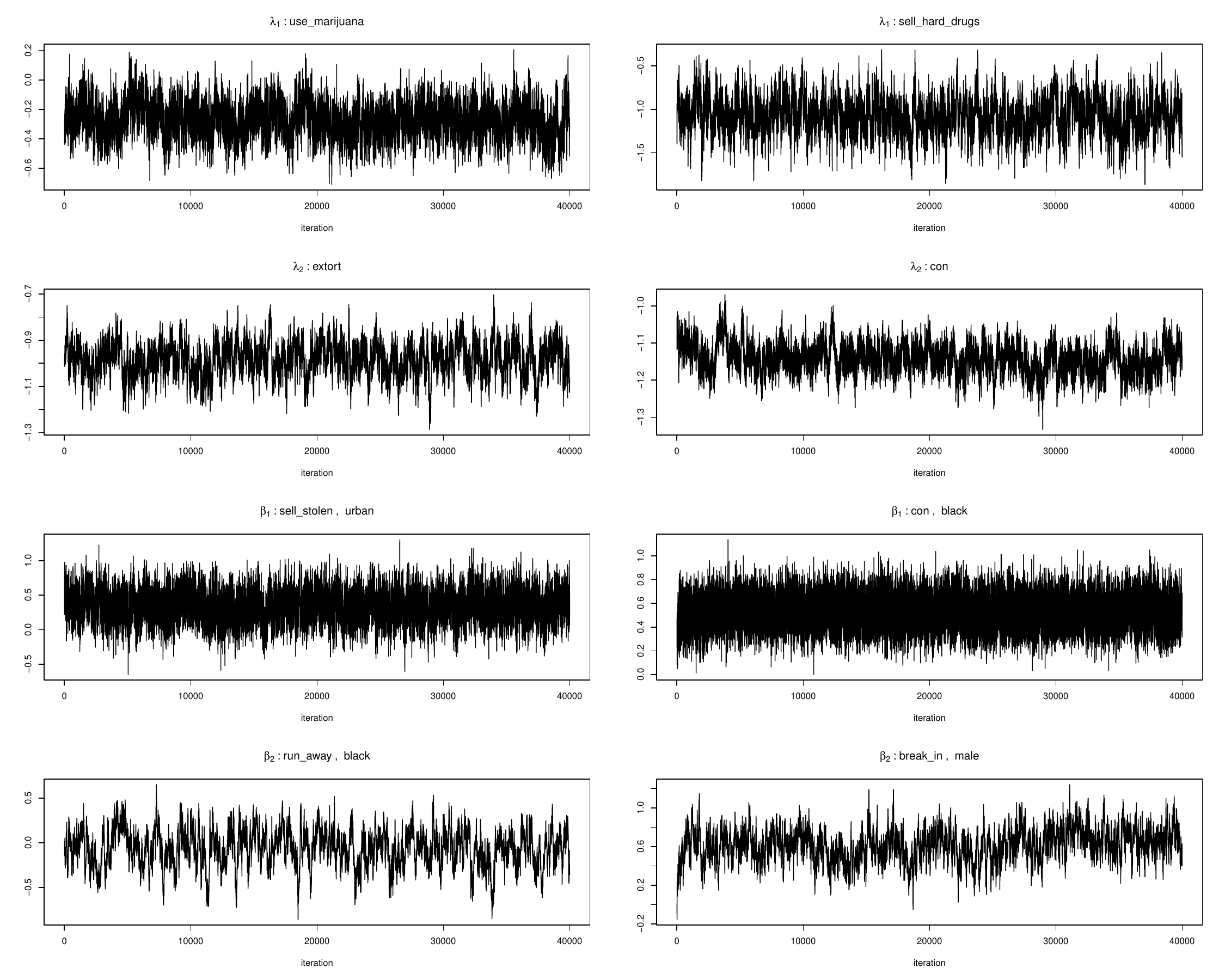}
    \caption{Trace plots of the Gibbs sampler for the selected parameters under GFZIP}
    \label{fig:traceplot}
\end{figure}

Figure~\ref{fig:traceplot} presents the trace plots of the Gibbs sampler for the selected parameters under GFZIP. 
For the factor loadings, the reordered series are shown. 
Although the model includes many latent variables in with the multiple hierarchy, it is seen the Gibbs sampler seems to be working reasonably well. 

Table~\ref{tab:real_lam} presents the posterior means and 95\% credible intervals for the factor loadings under GFZIP. 
Except for \texttt{fight} and \texttt{gambling} for at-risk, 
the 95\% credible intervals for all activities do not include zero. 
Among the credible at-risk loadings $\vlam_1$, 
the three factor loadings with the largest magnitudes in the posterior means are those for \texttt{sell\_hard\_drugs} ($-1.075$), \texttt{use\_hard\_drugs} ($-0.900$) and \texttt{sell\_marijuana} ($-0.594$), which are all drug-related loadings. 
For all Poisson factor loadings, $\vlam_2$, the 95\% credible intervals do not include zero. 
The loadings with the largest magnitudes in the posterior means are also the drug-related loadings such as \texttt{sell\_marijuana} ($-3.374$), \texttt{use\_marijuana} ($-2.991$) and \texttt{use\_hard\_drugs} ($-2.572$). 
Therefore, the single common latent factor included in the model is interpreted as the drug-related factor.

\begin{figure}[H]
    \centering
    \includegraphics[width=0.48\textwidth]{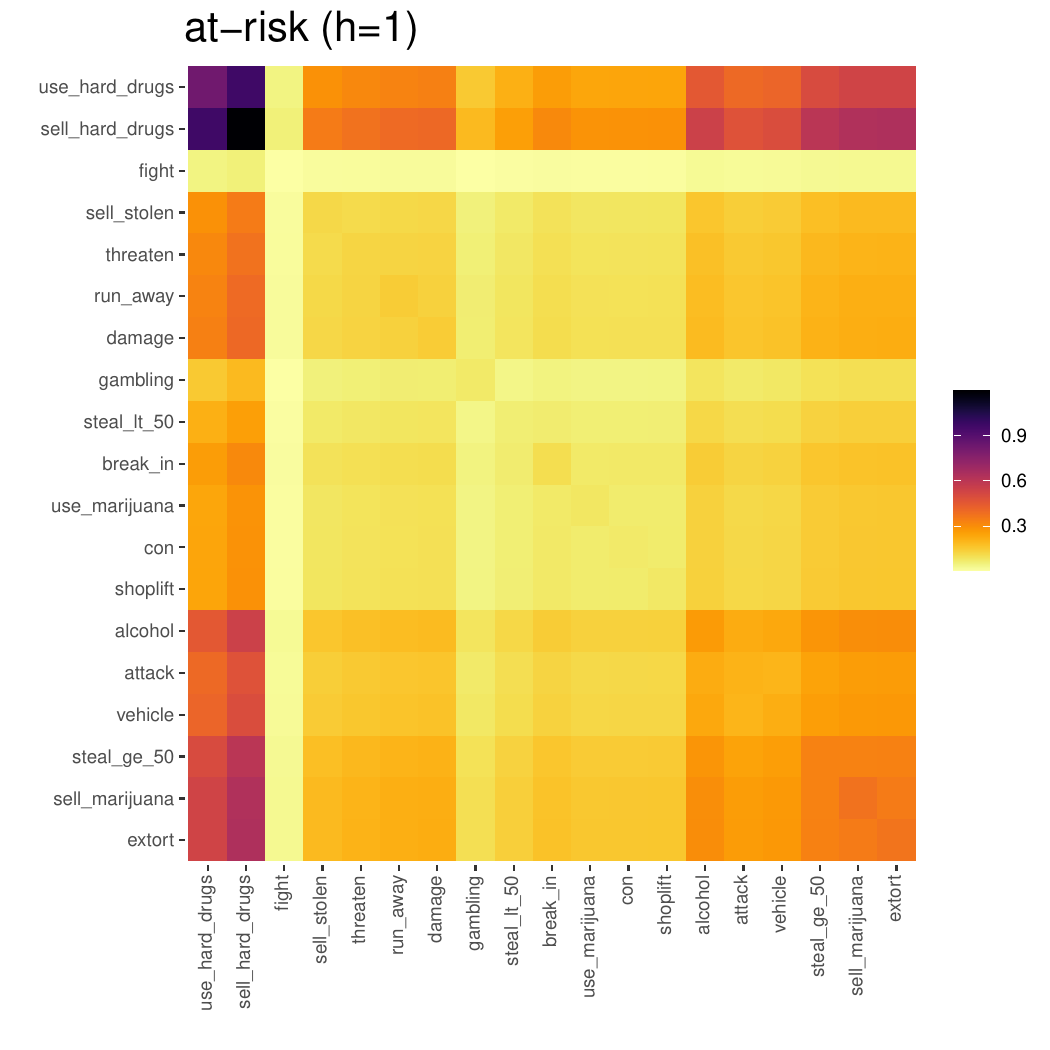}
    \includegraphics[width=0.48\textwidth]{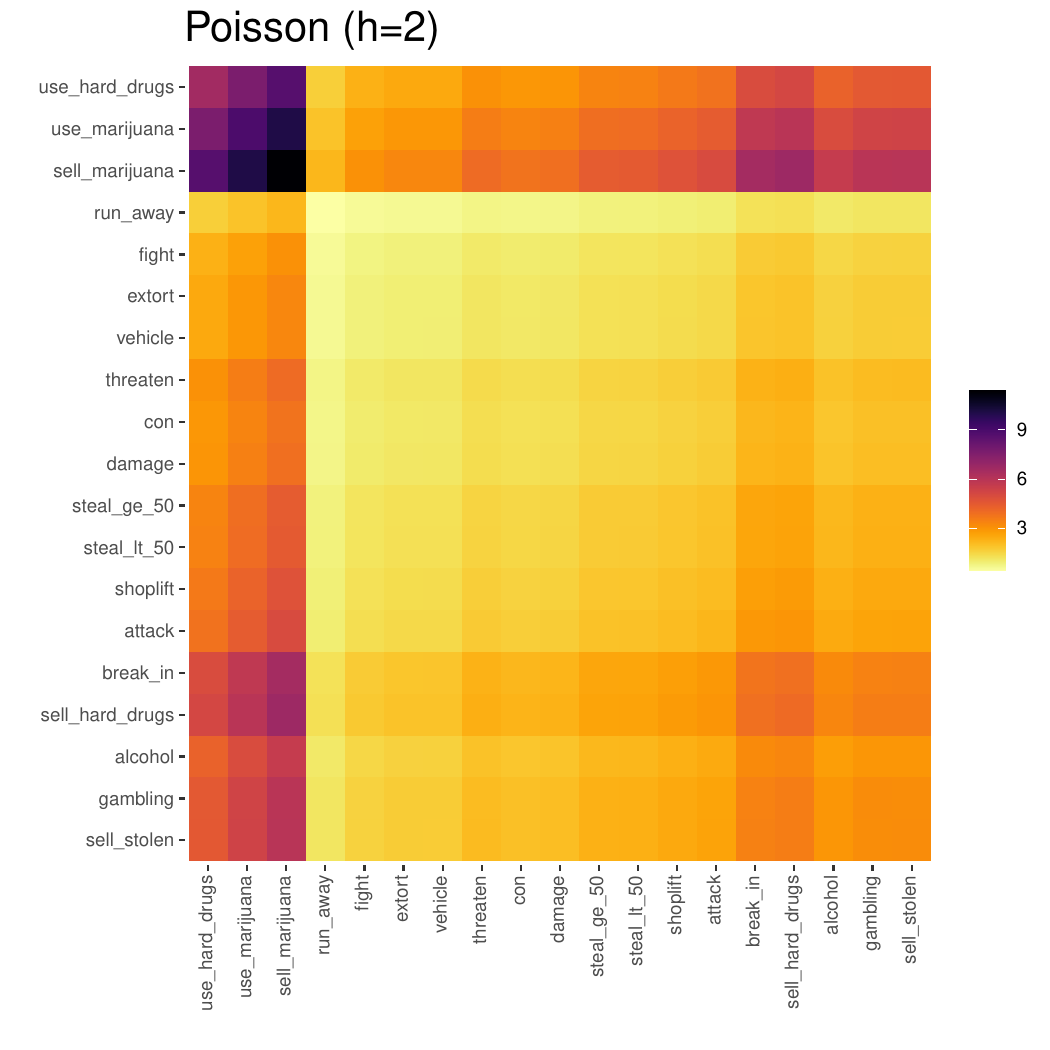}
    \caption{Posterior means of $\vlam_h\vlam_{h}'$ under GFZIP. The activities are ordered based on the hierarchical clustering.}
    \label{fig:lamlam}
\end{figure}

Figure~\ref{fig:lamlam} presents the heatmaps of the posterior means of $\vlam_h\vlam_h',\ h=1,2$ under GFZIP as indicators of the association within the at-risk and Poisson parts.  
The activities are ordered in each panel based on the hierarchical clustering for better visibility and interpretability. 
The darker the shades of the block for $\lambda_{hj}\lambda_{hj'}$, the greater the association between the activities $j$ and $j'$ in part $h$. 
In the top left corner of the left panel, there is a patch of noticeable dark shade. 
This part corresponds to the association among \texttt{sell\_hard\_drugs} and  \texttt{use\_hard\_drugs}, the two activities with the largest factor loadings in the at-risk part. 
The figures show that the involvement in these activities is also associated with the involvement in almost all the other activities except for \texttt{fight}, as indicated by the left and top edges of the heatmap. 
The activities such as \texttt{sell\_marijuana}, \texttt{alcohol} and \texttt{steal\_ge\_50} are relatively highly associated with \texttt{sell\_hard\_drugs} and  \texttt{use\_hard\_drugs}. 
These four activities also exhibit mild degree of association among themselves. 
In the top left corner of the right panel, there is also a dark patch indicating the association among \texttt{use\_harddrugs}, \texttt{use\_marijuana} and \texttt{sell\_marijuana}. 
Again, these activities exhibit association with all the other activities, as indicated by the darker bands along the left and top edges.

\begin{figure}[H]
    \centering
    \includegraphics[width=0.48\textwidth]{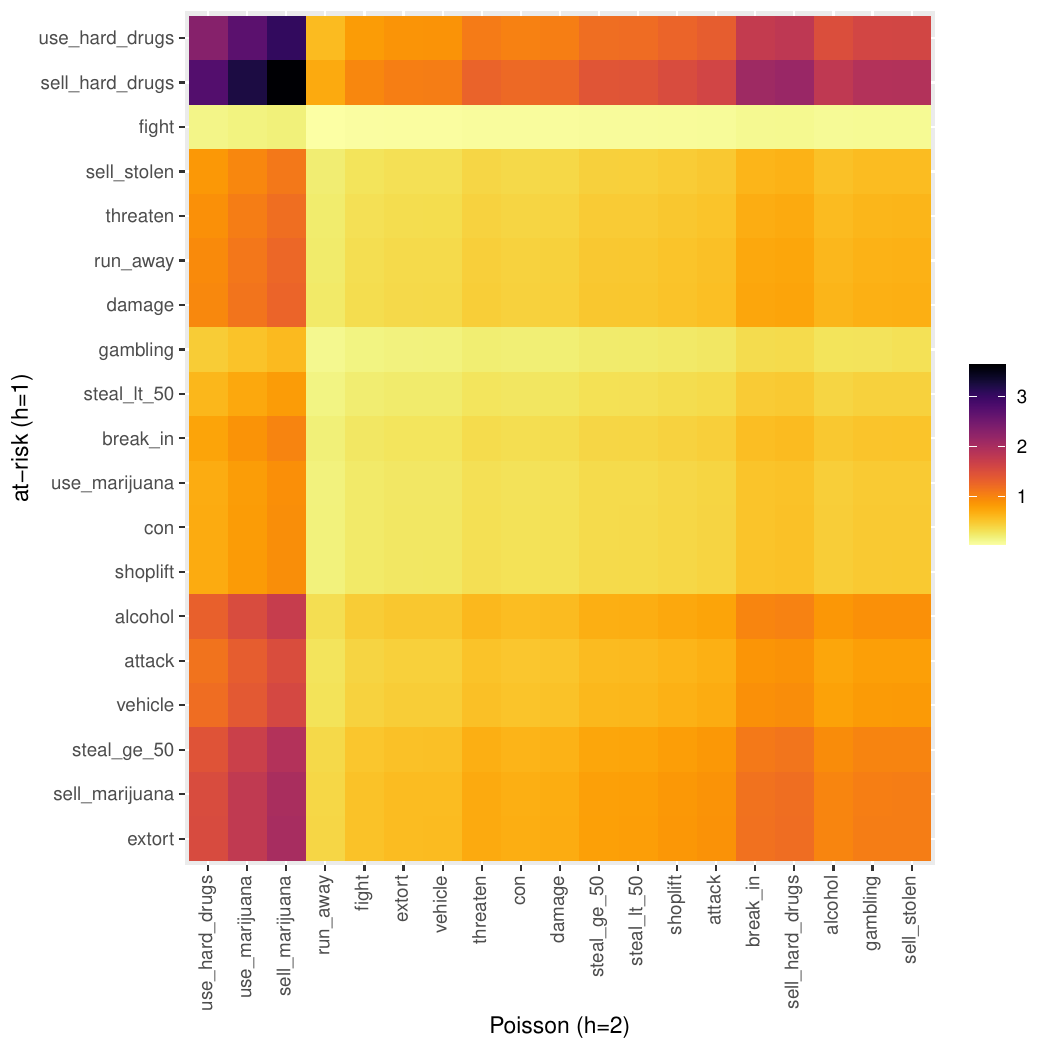}
    \caption{Posterior means of $\vlam_1\vlam_2'$ under GFZIP. The activities are ordered based on the hierarchical clustering.}
    \label{fig:lam12}
\end{figure}

Figure~\ref{fig:lam12} presents the heat map of the posterior means of $\vlam_1\vlam_2'$ representing the association between the at-risk and Poisson parts.  
The activities are ordered based on the hierarchical clustering. 
Similarly to Figure~\ref{fig:lamlam}, a dark patch for the drug-related activities between the at-risk and Poisson parts is recognisable.  
The figures show that being at-risk for \texttt{use\_hard\_drugs} and \texttt{sell\_hard\_drugs} is highly associated with the Poisson counts for themselves and \texttt{use\_marijuana}. 
It is also seen that being at-risk for using and selling hard drugs is also associated with the Poisson counts for all the other activities and that the Poisson counts for these three activities are also associated with being at-risk for most activities. 

\begin{figure}[H]
    \centering
    \includegraphics[width=0.48\textwidth]{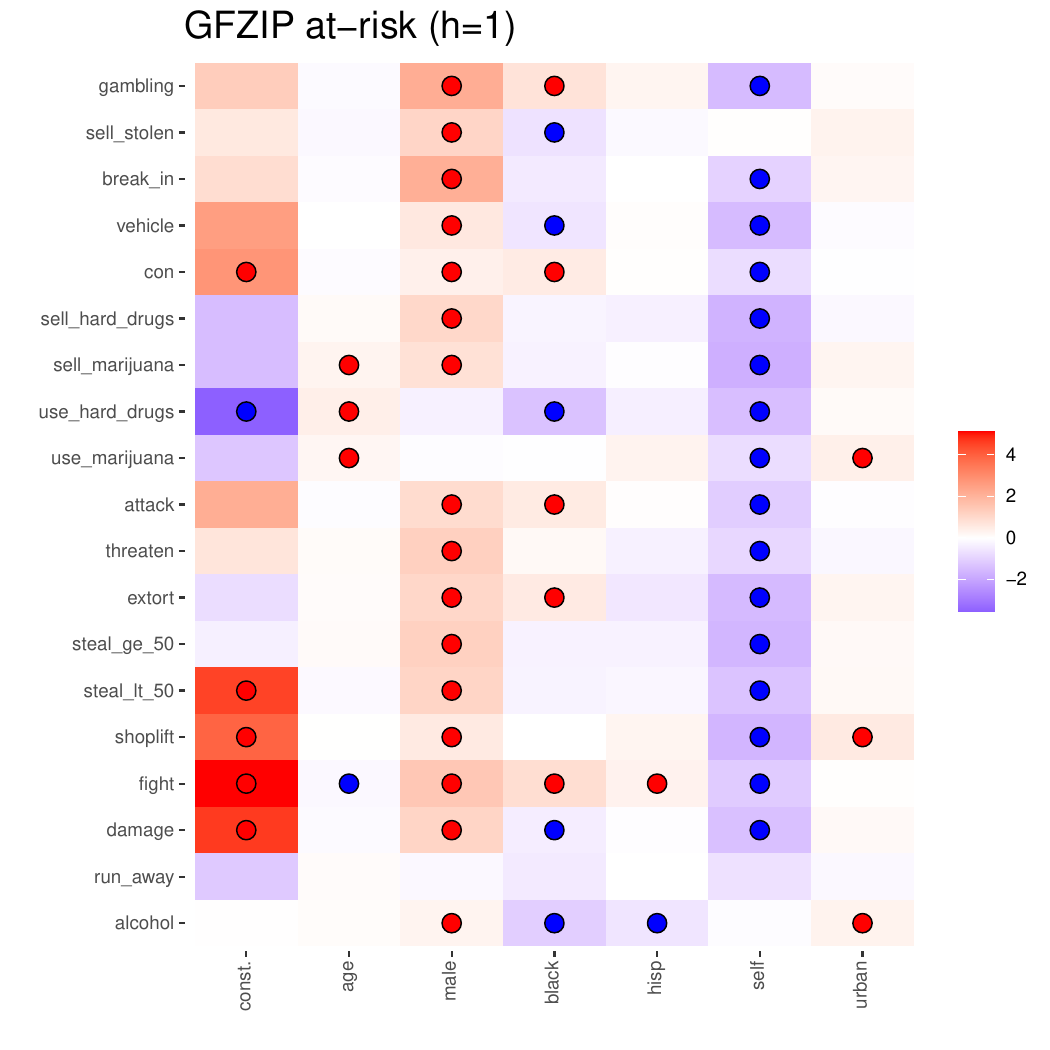}
    \includegraphics[width=0.48\textwidth]{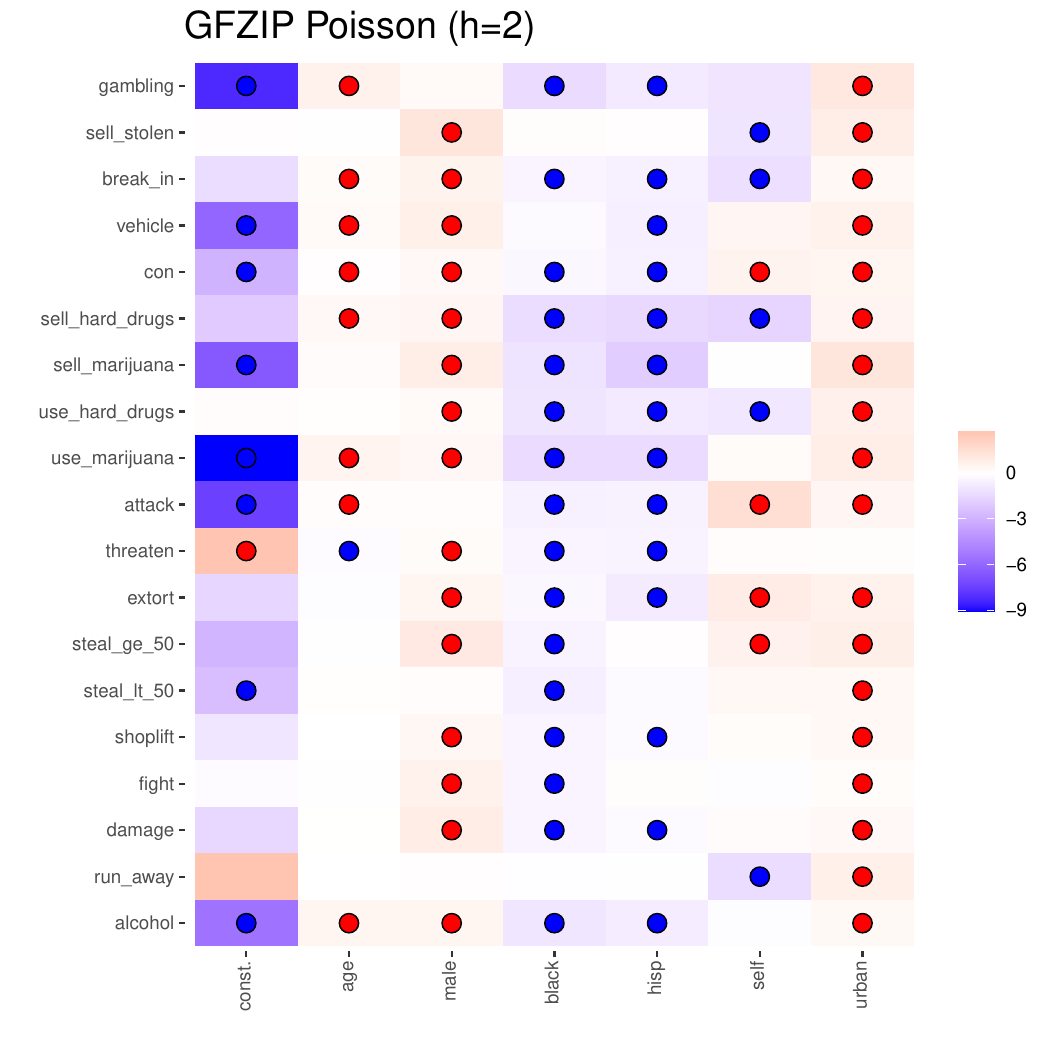}
    \caption{Posterior means of $\vbeta_h$ under GFZIP. The circles indicate the parameters for which the 95\% credible intervals do not include zero.}
    \label{fig:beta}
\end{figure}

Figure~\ref{fig:beta} presents the posterior means of $\vbeta_h$ for $h=1,2$ under GFZIP. 
The circles in the figure indicate the parameters for which the 95\% credible intervals do not include zero. 
Overall, the signs of the coefficients are the same for most activities. 
For example, the left panel shows \texttt{age} has positive effects on the at-risk probabilities for \texttt{sell\_marijuana}, \texttt{use\_hard\_drugs} and  \texttt{use\_marijuana}, but has negative effect on \texttt{fight}. 
\texttt{male} has positive effects on most activities other than \texttt{run\_away}, \texttt{use\_marijuana}, and \texttt{use\_hard\_drugs}. 
On the other hand, \texttt{self} has negative effects on the at-risk probabilities for most activities other than \texttt{alcohol}, \texttt{run\_away} and \texttt{sell\_stolen}. 
This is expected because the higher the self-esteem score, the less likely youths are to engage in illegal activities.

In the right panel, \texttt{urban} and \texttt{male} positively affect the Poisson counts of most activities. 
An urban environment would offer more opportunities for various types of illegal activities. 
Combined with the results on the at-risk coefficient for \texttt{male}, male youths are more likely to be involved in illegal activities, and their involvements are more frequent. 
\texttt{self} has a positive impact on the frequencies of the activities such as \texttt{attack}, \texttt{extort}, \texttt{steal\_ge\_50} and \texttt{con}.  
Most of these activities typically involve aggressive behaviour towards other individuals or audacity.
Therefore, higher self-esteem would increase the frequency of those activities.  
On the other hand, \texttt{self} has negative impacts on the frequencies of \texttt{sell\_stolen}, \texttt{break\_in}, \texttt{sell\_hard\_drugs} and \texttt{use\_hard\_drugs}, \texttt{run\_away}. 
It would be intuitive that the frequency of these activities, especially drug-related activities and running away, is associated with lower self-esteem.

\begin{figure}[H]
    \centering
    \includegraphics[width=0.6\textwidth]{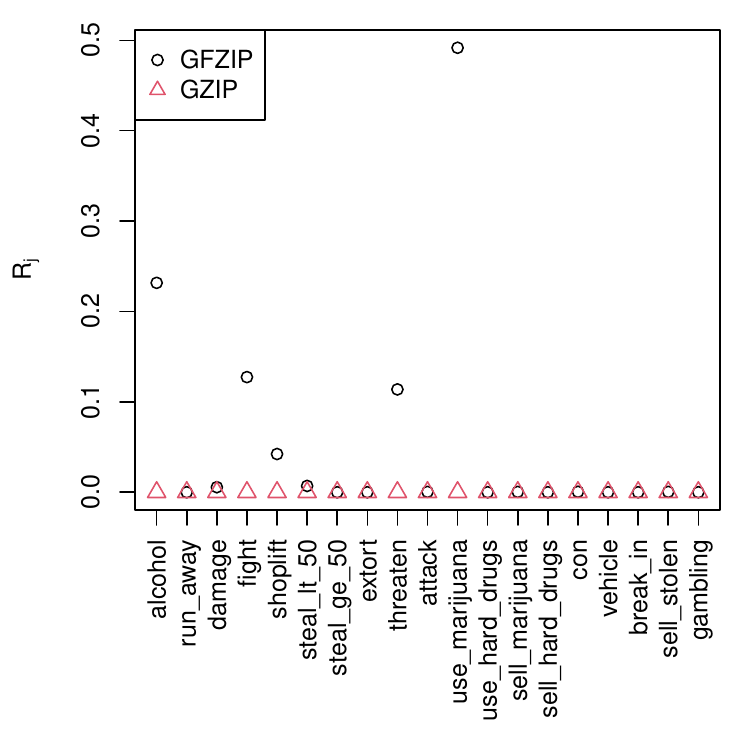}
    \caption{Proportions of at-risk youths among those who answered `never'}
    \label{fig:atrisk_prop}
\end{figure}

Finally, we estimate the proportions of at-risk youths among those who answered `never' for each activity based on \eqref{eqn:Rj}. 
These are the estimated fractions of youths involved in the activities, but their responses on the frequency of involvement happened to be zero one year before the interview. 
Figure~\ref{fig:atrisk_prop} presents $\hat{R}_j$ for 19 activities under GFZIP and GZIP. 
Under the proposed GFZIP, $\hat{R}_j$ for \texttt{alcohol}, \texttt{fight}, \texttt{threaten}  and \texttt{use\_marijuana} are above $0.1$. 
Among those activities, \texttt{use\_marijuana} resulted in the largest $\hat{R}_j$ of $0.492$. 
The result implies that nearly half of the youths who responded `never' actually are regular users but did not use them during the one year before the interview. 
$\hat{R}_j=0.232$ for \texttt{alcohol} is the second largest, followed by $0.127$ for \texttt{fight} and $0.114$ for \texttt{threaten}. 
These activities might be more common among youths, as indicated by non-zero proportions in Figure~\ref{fig:nlsy_hist}, compared to the other activities with a higher criminal nature, such as selling drugs and stealing vehicles. 
On the contrary, $\hat{R}_j$'s for the rest of the activities are zero or almost zero. 
The figure also shows that under GZIP $\hat{R}_j=0$ for all activities. 
The results for the activities such as \texttt{use\_marijuana} and \texttt{alcohol} are suspected to be false negative, as observed in the simulation study. 
The result under the proposed model is more reasonable and indicates the efficacy of leveraging shared information among activities through the latent factors.

\begin{table}[H]
    \caption{Posterior predictive loss for NLSY79 data}
    \label{tab:ppl}
    \centering
    \begin{tabular}{rrrrr}\toprule
        \multicolumn{3}{c}{GFZIP} & GZIP & GZINB \\\cmidrule(lr){1-3}
         $K=1$ &$K=2$&$K=3$\\\hline
         928.3 & 1900.3 & 3188.0 & 943.1 & 7360.4 \\
         \bottomrule
    \end{tabular}
\end{table}

\begin{table}[H]
    \centering
    \caption{Posterior means and 95\% credible intervals (CI) for the factor loadings under GFZIP}
    \label{tab:real_lam}
    \begin{tabular}{lrrrrrr}\toprule
 & \multicolumn{3}{c}{At-risk ($\vlam_1$)} & \multicolumn{3}{c}{Poisson ($\vlam_2$)}\\
 \cmidrule(lr){2-4}\cmidrule(lr){5-7}
 &Mean & \multicolumn{2}{c}{95\% CI} &Mean & \multicolumn{2}{c}{95\% CI}\\\hline
\texttt{alcohol}           & -0.511 &(-0.702, & -0.309) & -1.670 & (-1.744, & -1.595) \\
\texttt{run\_away}         & -0.372 &(-0.586, & -0.149) & -0.662 & (-0.857, & -0.493) \\
\texttt{damage}            & -0.379 &(-0.554, & -0.199) & -1.166 & (-1.262, & -1.076) \\
\texttt{fight}             & -0.056 &(-0.205, &  0.100) & -0.913 & (-0.975, & -0.853) \\
\texttt{shoplift}          & -0.275 &(-0.448, & -0.098) & -1.423 & (-1.510, & -1.339) \\
\texttt{steal\_lt\_\$50}   & -0.243 &(-0.429, & -0.054) & -1.340 & (-1.443, & -1.242) \\
\texttt{steal\_ge\_\$50}   & -0.564 &(-0.821, & -0.307) & -1.328 & (-1.495, & -1.177) \\
\texttt{extort}            & -0.598 &(-0.836, & -0.362) & -0.983 & (-1.131, & -0.848) \\
\texttt{threaten}          & -0.354 &(-0.498, & -0.207) & -1.198 & (-1.262, & -1.132) \\
\texttt{attack}            & -0.448 &(-0.655, & -0.238) & -1.498 & (-1.638, & -1.364) \\
\texttt{use\_marijuana}    & -0.271 &(-0.496, & -0.038) & -2.991 & (-3.146, & -2.838) \\
\texttt{use\_hard\_drugs}  & -0.900 &(-1.174, & -0.624) & -2.572 & (-2.792, & -2.375) \\
\texttt{sell\_marijuana}   & -0.594 &(-0.927, & -0.263) & -3.374 & (-3.648, & -3.028) \\
\texttt{sell\_hard\_drugs} & -1.075 &(-1.483, & -0.660) & -2.004 & (-2.391, & -1.699) \\
\texttt{con}               & -0.273 &(-0.411, & -0.136) & -1.141 & (-1.217, & -1.060) \\
\texttt{vehicle}           & -0.465 &(-0.644, & -0.287) & -0.986 & (-1.088, & -0.885) \\
\texttt{break\_in}         & -0.296 &(-0.587, &  0.006) & -1.949 & (-2.129, & -1.784) \\
\texttt{sell\_stolen}      & -0.329 &(-0.564, & -0.093) & -1.784 & (-1.928, & -1.641) \\
\texttt{gambling}          & -0.179 &(-0.602, &  0.269) & -1.777 & (-2.175, & -1.470) \\\bottomrule
\end{tabular}
\end{table}

\section{Conclusion}\label{sec:conc}
We have proposed the Poisson factor zero-inflated model for multiple grouped count data, which includes latent factors to account for association among the multiple count responses. 
Based on the data augmentation, P\'{o}lya-Gamma augmentation and parameter expansion, we have developed an efficient MCMC algorithm. 
The identification of the factor components is achieved through the post-processing algorithm. 
We have demonstrated the efficacy of the proposed model through the numerical examples. 
Notably, in the analysis of illegal activities of youths, we have found a single common factor, which can be interpreted as the drug-related factor, producing a strong association among the drug-related activities both in at-risk and Poisson parts. 
The proposed model also revealed the individuals at risk among those who reported zero in each activity, while treating each activity separately completely failed to do so.

\section*{Acknowledgement}
This work was supported by JSPS KAKENHI (\#21K01421, \#21H00699, \#20H00080, \#22K13376, \#24K00244).

\bibliographystyle{chicago}
\bibliography{Ref}

\end{document}